\documentclass[journal=jacsat,manuscript=article]{achemso}

\usepackage[version=3]{mhchem} % Formula subscripts using \ce{}
\usepackage[utf8]{inputenc}
\usepackage{amssymb}

\usepackage{soul}
\usepackage[dvipsnames]{xcolor}

\usepackage{chemformula}
\usepackage{siunitx}

\author{Rajam Elancheliyan}
\affiliation{Laboratoire Charles Coulomb, UMR 5221, CNRS–Universit\'{e} de Montpellier, F-34095 Montpellier, France}
\author{Giovanni Del Monte}
\affiliation{National Research Council–Institute for Complex Systems (CNR-ISC), Sapienza
University of Rome, 00185 Rome, Italy}
\altaffiliation{Department of Physics, Sapienza University of Rome, 00185 Rome, Italy}
\author{Edouard Chauveau}
\affiliation{Laboratoire Charles Coulomb, UMR 5221, CNRS–Universit\'{e} de Montpellier, F-34095 Montpellier, France}
\author{Simona Sennato}
\affiliation{National Research Council–Institute for Complex Systems (CNR-ISC), Sapienza University of Rome, 00185 Rome, Italy}
\altaffiliation{Department of Physics, Sapienza University of Rome, 00185 Rome, Italy}
\author{Emanuela Zaccarelli}
\affiliation{National Research Council–Institute for Complex Systems (CNR-ISC), Sapienza University of Rome, 00185 Rome, Italy}
\altaffiliation{Department of Physics, Sapienza University of Rome, 00185 Rome, Italy}
\email{emanuela.zaccarelli@cnr.it}
\phone{+39 (0)6 49913524}
\author{Domenico Truzzolillo}
\affiliation{Laboratoire Charles Coulomb, UMR 5221, CNRS–Universit\'{e} de Montpellier, F-34095 Montpellier, France}
\email{domenico.truzzolillo@umontpellier.fr}
\phone{+33 (0)467 143589}

\title[An \textsf{achemso} demo]
{The role of charge content in the two-step deswelling of Poly(N-isopropylacrylamide)-based microgels}

\abbreviations{PNIPAm, AAc, $f_{AAc}$, SLS, DLS }
\SectionNumbersOn
\begin{document}

%\begin{tocentry}

%Some journals require a graphical entry for the Table of Contents.
%This should be laid out ``print ready'' so that the sizing of the

%Inside the \texttt{tocentry} environment, the font used is Helvetica
%8\,pt, as required by \emph{Journal of the American Chemical

%The surrounding frame is 9\,cm by 3.5\,cm, which is the maximum
%permitted for  \emph{Journal of the American Chemical Society}
%graphical table of content entries. The box will not resize if the
%content is too big: instead it will overflow the edge of the box.

%This box and the associated title will always be printed on a
%separate page at the end of the document.

%\end{tocentry}

\begin{abstract}
 Poly(N-isopropylacrylamide)-based microgels are soft colloids undergoing a Volume Phase Transition (VPT) close to ambient temperature. Although widely employed for fundamental research and application purposes, the modifications of the microgel internal structure occurring at the VPT are not yet completely understood, especially concerning the role of electrostatics. Here we study in detail, both experimentally and numerically, the effect of the addition of acrylic acid (AAc) co-monomer on the microgel deswelling process. By combining viscosimetry, light scattering and electrophoresis, we show that the progressive addition of AAc increases the microgel mass and suppresses the occurrence of the VPT, progressively shifting  the microgel collapse to higher temperatures. Most importantly, it also highly enhances the two-step deswelling of these submicron-sized networks, so that the inner core collapses at temperatures always lower than those marking the transition of the outer corona. These results indicate that a net increase of the charge density mismatch between the bulk and the surface of the microgels takes place. Numerical simulations fully confirm this scenario and clarify the impact of the charge distribution on the two-step deswelling, with mobile counterions efficiently screening the charges within the inner core, while leaving more monomers ionized on the surface. Our work unambiguously shows how electrostatic interactions influence the behavior of thermosensitive microgels in aqueous environment close to the VPT.        

\end{abstract}

\section{Introduction}

Microgels are soft colloidal particles consisting of crosslinked polymer networks whose size ranges from few tens of nanometers up to few microns. They are widely studied for a plethora of purposes spanning many research fields, such as biomedicine \cite{lima_ocular_2016, garcia-pinel_magnetically_2020}, sensors\cite{islam_poly_2014} and 3D bioprinting\cite{highley_jammed_2019}, and they are currently used in many industrial applications as biocompatible additives\cite{compaan_cross-linkable_2020}, delivery vehicles \cite{murthy_macromolecular_2003} and color-changing systems \cite{debord_color-tunable_2002}.
Stimuli-responsive microgels have gathered much attention due to the fact that they display a phase change or an abrupt single-particle volume change in response to external stimuli, such as a temperature, pH and/or ionic strength variation \cite{keidel_time-resolved_2018,debord_synthesis_2003, wypysek_tailoring_2020}, and many applications have been based on their tunability\cite{garcia-pinel_magnetically_2020, murthy_macromolecular_2003-1,nagayama_catalytic_1995, bergbreiter_poly_1998}. %Their intrinsic tunability makes them highly suitable for many potential applications ranging from drug and vaccine delivery \cite{garcia-pinel_magnetically_2020, murthy_macromolecular_2003-1} to smart sensors for various analytes \cite{jia_orderdisorder_2018}, and catalysis \cite{nagayama_catalytic_1995, bergbreiter_poly_1998}.
 
Among polymeric microgels, those made of poly(N-isopropylacrylamide) (PNIPAm) are particularly interesting, since they undergo a Volume Phase Transition (VPT) at ambient temperature: below a characteristic temperature, $T_c$ $\approx 32^{\circ}C$, these colloidal networks exist in a swollen, fully hydrated state, while above $T_c$ they collapse due to their increased hydrophobicity. %It has been demonstrated that the swelling/deswelling transition of PNIPAm microgels can be controlled either by temperature, ionic strength or pH \cite{del_monte_two-step_2021,wypysek_tailoring_2020}. 
PNIPAm microgels can be synthesized using standard emulsion polymerization in aqueous media\cite{pelton_preparation_1986} and their size, shape and structure have been investigated using various methods including x-ray, light and neutron scattering \cite{seelenmeyer_analysis_2000, seelenmeyer_small-angle_2001, stieger_small-angle_2004,sennato_double-faced_2021}, and electron or atomic force microscopy \cite{sennato_double-faced_2021}. The combination of many different experimental techniques to investigate the VPT of PNIPAm microgels made it possible to grasp a lot of information on this phenomenon, which is one of the most studied in soft condensed matter nowadays. 

The transition temperature as well as the physical nature of the volume phase transition can be modified by additives \cite{ishidao_swelling_1993,kim_saccharide_1995}, like salts \cite{park_sodium_1993} or surfactants \cite{zhang_super-absorbency_1992}, through changes in the solvent composition either by chemical insertion \cite{inomata_swelling_1995}, e.g. by embedding charged groups within the networks\cite{shibayama_dynamic_1996,kratz_influence_2000,karanastasis_one-pot_2020}, or by physical interpenetration of different polymer species \cite{nigro_chemical-physical_2021}. Despite the intense research effort, one basic aspect of the volume phase transition of microgels in aqueous media, namely the influence of electrostatic interactions, still remains relatively unexplored and poorly understood. 

Very recently, some of us \cite{del_monte_two-step_2021} have shown that crosslinked networks of PNIPAm homopolymers undergo a two-step volume phase transition, where the residual charges carried by the ionic initiators, mainly located in the outer corona, play a pivotal role. During an ascending temperature ramp, PNIPAm microgels deswell with their initiator-free core collapsing first, namely at temperatures lower than the initiator-rich corona. The fingerprint of such two-step deswelling has been identified in the onset of a minimum of the ratio $R_g/R_H$ between the gyration and the hydrodynamic radius of the microgels as a function of temperature \cite{del_monte_two-step_2021}. Simulations have further shown that this feature does not take place in the absence of charges\cite{del_monte_two-step_2021}.  

Previous works have also demonstrated that the properties of the VPT, in particular the critical temperature $T_c$, can be altered by incorporating a charged co-monomer such as Acrylic acid (AAc) \cite{snowden_colloidal_1996} into the PNIPAm network. In the presence of acrylic acid and for pH$>$pKa$^{(AAc)}$, the swelling/deswelling behaviour of microgels depends on the subtle balance between hydrophobic attraction and electrostatic repulsion due to the carboxylate groups borne by the charged co-monomers. Introducing electric charge by co-polymerization hence offers the possibility to embed these charges within the microgel and not simply at one chain end, as for the case of the ionic groups carried by the initiator. 

AAc units can be incorporated in microgels following three different synthesis strategies:
i) Poly(N-isopropylacrylamide-co-acrylic acid) microgels can be synthesized using emulsion polymerization technique \cite{kratz_influence_2000,huo_temperatureph_2006,burmistrova_effect_2011}, giving rise \emph{a priori} to random co-polymerization; ii) random distributions of AAc units can also be obtained by incorporating AAc monomers into PNIPAm networks in the form of cross-linker rather than simple co-monomers\cite{clara-rahola_structural_2012}; iii) polymer interpenetration—IPN (Interpenetrated Polymer Network) can also be employed to produce pH responsive PNIPAm/PAAc microgels that have been widely investigated\cite{mattsson_soft_2009,nigro_chemical-physical_2021}. %both at the single particle level \cite{nigro_chemical-physical_2021} and in concentrated suspensions \cite{mattsson_soft_2009}. 
Among these methods, random co-polymerization (i) can be considered as the natural extension of the more standard radical polymerization of homopolymers towards the formation of chemically heterogeneous polymeric colloids or polymers. Indeed, this route does not assign any particular structural feature to the charged units, either in terms of topology or of connectivity, with respect to the neutral monomers. However, despite this strategy has been widely used to tune the size and the responsiveness of microgels, its impact on the microgel structure across the VPT remains currently unknown.

The aim of the present work is to investigate the role played by the AAc fraction on the two-step deswelling of Poly(N-isopropylacrylamide-co-acrylic acid) microgels. 
By means of viscosimetry, light scattering and electrophoresis, we show that the presence of AAc monomers in the polymer structure increases the microgel mass and suppresses the VPT, progressively shifting  the microgel collapse to higher temperatures. In addition, it also largely enhances the two-step deswelling phenomenon. In order to isolate the role played by the charge content on the volume phase transition of the PNIPAm-co-AAc microgels, we have studied microgel suspensions in the absence of added salts, acids and bases, with all carboxylic groups borne by the acrylic acid monomers in their deprotonated state.
In this regime a unique continuous deswelling for ascending temperature ramps is observed in both the hydrodynamic size and the gyration radius of the particles in agreement with other syntheses  \cite{kratz_influence_2000,karanastasis_one-pot_2020}.       
To have a closer look into the mechanisms driving such a phenomenon we combine the experiments with state-of-the-art computer simulations using a coarse-grained bead-spring model, based on a realistic representation of the polymer network at a monomer-resolved level. Simulations further shed light on the role played by the total charge and its spatial distribution around the VPT, corroborating the enhancement of the two-step deswelling observed in experiments.

The remainder of this paper is organized as follows. In section \ref{matmeth} we give details of the synthesis, the experimental techniques and the numerical methods used to investigate single microgel structure and their VPT. In section \ref{results} we report both experimental and numerical results. Finally, in Section \ref{conclusion} we make some concluding remarks and summarize the key results of the work. 

\section{Materials and Methods}\label{matmeth}

\subsection{Synthesis}

Poly(N-isopropylacrylamide-co-acrylic acid) microgels were synthesized using emulsion polymerization\cite{senff_temperature_1999}. 150 mL of ultra-pure water was introduced in a 250mL three-necked flask, and degassing was carried out using vacuum/argon cycles. Vacuum was achieved by a vane pump and the argon/vacuum sequence was repeated 6 times. Finally, argon was bubbled for 15 min. After having completed degassing, 3 to 4 mL of the (degassed) water was withdrawn via a syringe to dissolve the initiator (potassium peroxodisulfate, KPS – purchased from Sigma Aldrich and used without further purification) that was added in a later stage.  Once the bubbling was stopped and the mechanical agitation was set up (via Teflon rotating anchor), the two side necks were plugged.
Sodium Dodecyl Sulfate (SDS – purchased from Sigma Aldrich and used without further purification) was added in the flask right before the solution was heated up to the desired temperature ($T_s=70\pm 1$ $^{\circ}$C). SDS/NIPAm molar ratio has been fixed to 0.0085. The largest relative variation of the SDS mass added during the different syntheses amounts to 5\%. A relative increase of the swollen microgel size \cite{misra_influence_2020} that does not exceed 4\%, was expected due to such differences and is barely detectable. 

The target quantities of i) N-isopropylacrylamide (NIPAm) (from Sigma Aldrich, used without further purification), ii) N,N-methylene-bis-acrylamide (BIS) (from Sigma Aldrich, used without further purification) and iii) acrylic acid (AAc) (from Sigma Aldrich and used without further purification) were introduced in turn as reported in table (Table 1 in SM). The AAc mass content has been chosen in order to vary its monomer fraction, namely the ratio between the number of AAc monomers and the total number of monomers participating to the radical polymerization, from 0\% to 5\% (Coded as AC0, AC1, AC3 and AC5 in Table \ref{tbl:Summary}).   
During the heating ramp, the initiator (KPS) was dissolved in 3-4 mL of deionized degassed water and injected slowly once the temperature of the batch reached $T_s$. The mixture was left under stirring at $T_s$ for 6 hours. The polymerization terminated spontaneously. %Though we did not monitor the pH of the suspensions right after the synthesis their nominal pH, calculated based on the amount of added AAc, ranges from 3.52 (5\% AAc) to 3.87 (1\% AAc sample), at which we expect that the high protonation degree of acrylic monomers increases their solubility into the NIPAm droplets formed at 70 $^\circ$C and homogenize their distribution within the emerging network.
All samples have been purified via three consecutive centrifugation/supernatant removal cycles \cite{conley_relationship_2019, del_monte_two-step_2021} and 19.5 mg (2 mM) of Sodium Azide have been added to prevent bacterial growth. 
Finally we point out that the pH of all samples prepared for viscosimetry, light scattering and electrophoresis experiments was never less than 6. At these pH values, deprotonation of the carboxylic groups carried by AAC monomers occurs, resulting in a net electrostatic repulsion between the negatively charged carboxylate residues (pKa$^{(AAc)}\simeq$4.25)\cite{snowden_colloidal_1996}. 
On the contrary, the nominal pH of the samples during the synthesis and well before the complete polymerization, that can be calculated on the basis of the amount of the added AAc monomers, ranges from 3.52 (5\% AAc) to 3.87 (1\% AAc), at which the acidic monomers are mainly protonated (neutral). 

\subsection{Viscosimetry}
We performed viscosimetry experiments to measure the colloidal volume fraction of the microgels and to discern the effect of acrylic acid addition on the mass of a single microgel, as detailed in section \ref{results}. 

The colloidal volume fraction is defined as $\varphi$ = $n_pv_p$, where $n_p$ is the number density of colloids and $v_p = \pi d^3/6$ the volume of a single particle of diameter $d$ at infinite dilution.  Experimentally, only the concentration $c$ (mg/ml) of a (purified) suspension can be directly measured, by weighting a small aliquot of the sample before and after removing the solvent by evaporation.
Since the microgels are highly swollen, their mass density is essentially the same as that of the solvent; consequently, $c$ and $\varphi$ are proportional, $\varphi = kc$. We determined the constant $k$ matching the $c$ dependence of the zero-shear viscosity of the purified suspensions $\eta$ with the values expected from the Einstein equation\cite{truzzolillo_overcharging_2018},

\begin{equation}
   \frac{\eta}{\eta_0} = 1 + 2.5kc,
    \label{eq:Einstein}
\end{equation}
where $\eta_0$ is the viscosity of the solvent. By fitting $\eta/\eta_0$ to a straight line, we obtained $k$.

Such a procedure has been repeated for all samples. The measurements were performed at $T = 20$ $^{\circ}$C using an Anton Paar Lovis 2000 ME micro-viscosimeter in the range 0.02 $mg/ml$ $< c <$ 1.6 $mg/ml$, where $\eta$ increases linearly with $c$. To double-check that the mutual interactions between microgels did not affect viscosity in this concentrantion range we calculated a posteriori the ratio $d_c/\lambda_D$ between the center-to-center distance $d_c$ between microgels and the Debye length $\lambda_D$ of the suspensions taking into consideration only ions coming from the dissociated sodium azide. We obtained always $d_c/\lambda_D>100$, confirming that the suspended particles interact only rarely due to collisions and their mutual interaction do not participate to the increased viscosity.    

\subsection{Light scattering}

Light scattering measurements were performed to study the structure of PNIPAm-AAc microgels as a function of temperature. A laser source ($\lambda$ = 532 nm) was used for all the scattering measurements that were performed via an Amtec-goniometer allowing to collect the light at scattering angles in the range $16^{\circ}\leq\theta\leq 150^{\circ}$ corresponding to scattering wave-vectors $4.4$ $\mu m^{-1}$ $\le q \le 30.3$ $\mu m^{-1}$. All scattering experiments have been performed in dilute samples: an aliquot of each mother batch was diluted 50 times in deionized water, which resulted in a final microgel volume fraction $\varphi\le 0.5\%$. The pH of the samples at 20 $^{\circ}$C ranged from 6.2 (5$\%$AAc) to 6.9 (0\% AAc). In such a small slightly acidic pH interval the size of PNIPAm-co-AAc micro- and macrogels is not remarkably affected by pH variation \cite{pei_effect_2004,al-manasir_effects_2009}.
Before each data acquisition, the samples were thermostatted for 10 mins via a recirculating bath with 0.1 $^{\circ}$C accuracy. 

The hydrodynamic radius, $R_H$, and the size distribution of the microgels were characterized by means of dynamic light scattering. The scattered light intensity was collected at a fixed scattering angle ($\theta = 70^{\circ}$) correspondent to a scattering vector $q_{DLS}=18$ $\mu m^{-1}$ and analyzed using a digital autocorrelator. The time decay of the autocorrelation function $F_s(\vec{q},t)^2$ was then fitted by a second-order cumulant expansion to extract the diffusion coefficient $D$ as shown below \cite{frisken_revisiting_2001}:
\begin{equation}
   F_s(\vec{q}_{DLS},t)^2 \propto exp\left(-2q^2_{DLS}Dt\right) \left[1+\frac{\mu _2t^2}{2!}+ o(t^3)\right]^2
    \label{eq:cumulant}
\end{equation}

where $\mu_2$ is related to the second moment of the distribution of the diffusion coefficients of the suspended particles. The average diffusion coefficient is then used to estimate the hydrodynamic radius using the Stokes-Einstein relation: $D = K_BT/6 \pi \eta_o R_H $, where $K_B$ is the Boltzmann constant, $T$ is the bath temperature and $\eta_0$ is the zero-shear viscosity of the solvent. The corresponding size dispersion and polydispersity index are respectively $\sigma_{R_H} = \sqrt{\mu_2}R_H/(Dq^{2}_{DLS})$ and $\gamma=\mu_2/D^2q_{DLS}^4$. The polydispersity index for all the samples investigated here never exceeded 0.23 (See SI).  

The gyration radius, $R_g$, was measured by collecting the intensity of the light $I(q)$ scattered by the microgel samples at different scattering angles. The scattered light was subsequently fitted to the Guinier equation\cite{guinier_small-angle_1955} to extract $R_g$:

\begin{equation}
   I(q) = I(0) \exp\left[-\frac{(qR_g)^2}{3}\right],
    \label{eq:guiner}
\end{equation}

where $I(0)$ is a constant depending on the number of particles in the scattering volume and on the scattering factor of a single particle. The Guinier regime for all samples was attained in the range $0.5 \le qR_g \le 2.5$, coherently with previously reported microgel syntheses \cite{gasser_form_2014, clara-rahola_structural_2012}. The uncertainty on $R_g$ is given by the fit error, the latter being less than $2.5\%$ of the best-fit value. For the sample with the largest acrylic acid content (AC5 hereafter) most of the data points lied in range $1 \le qR_g \le 2.5$, hence we employed an extended Guinier approximation as reported in \cite{mohammed_dilute_2019}.
\begin{equation}
   I(q) = I(0) \exp\left[-\frac{(qR_g)^2}{3}-Cq^4\right].
    \label{eq:guiner-ext}
\end{equation}
where the constant $C$ quantifies the $O(q^4)$ term in the expansion of the form factor of the microgels around $q=0$. In all cases this term resulted negligible, since $R_g$ values extracted via equation \ref{eq:guiner-ext} deviated less than 3\% from the values extracted by employing equation \ref{eq:guiner}.     

\subsection{Electrophoresis}
The electrophoretic mobility of the microgels has been measured by means of a NanoZetaSizer apparatus (MALVERN), that is integrated with a laser Doppler electrophoresis technique. The average electrophoretic mobility is determined using the Phase Analysis Light Scattering (PALS) method \cite{tscharnuter_mobility_2001}, which is useful especially at high ionic strengths, where mobilities are low. In these cases the setup in the PALS configuration can measure mobilities two orders of magnitudes lower than traditional light scattering methods based on the shifted frequency spectrum (spectral analysis). We measured the electrophoretic mobility at 20 $^{\circ}$C and 60 $^{\circ}$C after a proper thermalization to monitor the effect of acrylic acid addition onto the mobility of swollen and collapsed microgels.

\subsection{\label{subsec:simulations}Numerical simulations}
Numerical simulations have been performed to unveil the microscopic mechanisms underlying the structural swelling-deswelling transition in PNIPAm-co-AAc microgels.
We rely on a coarse-grained bead-spring model\cite{grest1986molecular} of the polymer network composing the particles, wherein single polymer chains are modelled as sequences of beads representing single uncorrelated polymer segments.
Beads interact among each other through a repulsive steric interaction modelled with the Weeks-Chandler-Anderson (WCA) potential:

\begin{equation}
\label{eq:wca}
V_{\text{WCA}}(r)  =  
\begin{cases}
4\epsilon\left[\left(\frac{\sigma}{r}\right)^{12}-\left(\frac{\sigma}{r}\right)^6\right]+\epsilon & \quad \text{if} \quad r \le 2^{1/6}\sigma  \\
0 & \quad \text{if}  \quad  r > 2^{1/6}\sigma
\end{cases}
\end{equation}
where $\epsilon$ and $\sigma$ are the energy and length units, respectively.
Connected beads also interact via  the finitely extensible nonlinear elastic potential (FENE):
\begin{equation}
V_{\text{FENE}}(r)  = 
-\epsilon k_F{R_0}^2\log\left[1-{\left(\frac{r}{R_0\sigma}\right)}^2\right], \quad r < R_0\sigma 
\end{equation}
where $R_0=1.5\sigma$ is the maximum bond distance and $k_F=15$ is a stiffness parameter influencing the rigidity of the bond. The values of $R_0$ and $k_F$ have been chosen in order to preserve the topology of the network during the simulation. These bonds cannot break during the simulation, mimicking strong covalent bonding.
Beads that are linked via FENE potential to two neighbours represent polymer segments containing NIPAm or AAc monomers only, while those mimicking crosslinkers  have fourfold valence.

Our simulations focus on the study of single microgel particles of $N\sim42000$ beads, obtained through a recently developed protocol~\cite{gnan2017in} able to generate fully-bonded, disordered polymer networks. In order to accurately describe the internal structure of microgels, we adopt an appropriate tuning of the radial distribution of the crosslinkers, as established in previous work~\cite{ninarello2019modeling} through a quantitative comparison of the microgel form factors across the VPT. In this way, we ensure that the simulated networks have a realistic core-corona structure similar to the experimental one. In this work, we analyse for each simulated conditions a single microgel topology having a crosslinker fraction similar to the experimental one, i.e. $c = $5.0\%.

Within the model we also include electrostatic interactions to appropriately describe the presence of ions in the polymer backbone stemming both from dissociated AAc monomers and from \ch{SO4^-} molecular groups, that result from the thermal splitting of the persulfate ions \ch{S2O8^2-} (stemming from the solvation of KPS) and initiate free-radical polymerisation. To model the presence of these ionic species, we consider a fraction $f$ of the beads as having a negative charge and insert an equivalent number of counterions with positive charge to preserve the overall electro-neutrality. Following our previous works\cite{del2019numerical,del2020charge}, counterions sterically interact among each other and with microgel beads through the WCA potential, and their diameter is set to $\sigma_c=0.1\sigma$, while electrostatic interactions are given by the Coulomb potential:
\begin{equation}\label{eq:coulomb}
	V_{\rm coul}(r_{ij}) = \frac{q_i q_j \sigma}{e^{*2} r_{ij} } \epsilon
\end{equation}
where $q_i$ and $q_j$ are the charges of the beads, and $e^* = \sqrt{4\pi\varepsilon_0\varepsilon_r\sigma\epsilon}$ is its reduced unit, embedding the vacuum and relative dielectric constants, $\varepsilon_0$ and $\varepsilon_r$. $q_i$ and $q_j$ are set to the values $-e^*$ or $+e^*$ whether it refers to charged beads or counterions, respectively.

To match with the experimental systems, we study different charge fractions $f$ and distributions. In particular, we consider that charged groups coming from the initiator are distributed within the corona of the microgels~\cite{scotti_hollow_2018,del_monte_two-step_2021}, so that we assign a fraction of charged beads $f_s=3.2\%$, representing initiator molecules, to be located at a distance from the microgel center of mass greater than its radius of gyration.  On the other hand, in principle, we do not know where the AAc monomers are located. Therefore, we first consider that they can be found anywhere in the network and we distribute a varying fraction of them,  $f_r= 0 \%, 1 \%, 3 \%, 4 \%$, mimicking the experimental values, randomly throughout the microgel. However, since  part of the AAc monomers could be also preferentially located in the microgel periphery to minimize their mutual electrostatic repulsion during the polymerization process,
%\gdm{(taking place at pH conditions in which AAc monomers are deprotonated) but this is in contrast with what said in the introduction, that during the synthesis AAc monomers are protonated\ldots} \DT{I clarified further this point in Methods}, 
we additionally investigate the case with $f = f_s = 7.2\%$, where all charges (both coming from initiator and from AAc monomers) are located on the surface as the most extreme condition. This case will be directly compared with that having $f_s = 3.2\%$ and $f_r = 4\%$, which should be in closer agreement with the experimental synthesis conditions discussed in the Methods, in order to assess the effect of charges distribution on the swelling properties.

In all simulations crosslinkers are always set to be neutral and the solvent is implicitly taken into account through an effective potential which mimics the change in the polymer-solvent affinity by raising temperature:
\begin{equation}\label{eq:valpha}
V_{\alpha}(r)  =  
\begin{cases}
-\epsilon\alpha & \text{if } r \le 2^{1/6}\sigma  \\
\frac{1}{2}\alpha\epsilon\left\{\cos\left[\gamma{\left(\frac{r}{\sigma}\right)}^2+\beta\right]-1\right\} & \text{if } 2^{1/6}\sigma < r \le R_0\sigma  \\
0 & \text{if } r > R_0\sigma
\end{cases}
\end{equation}
where $\alpha$ is the solvophobicity parameter representing the effective temperature, which is varied from $\alpha=0$ (good solvent conditions) to $\alpha = 1.40$ (bad solvent conditions) to reproduce the swelling curve from the swollen state to the most collapsed ones~\cite{gnan2017in}. The constants $\gamma = \pi\left(\frac{9}{4}-2^{1/3}\right)^{-1}$ and $\beta = 2\pi - \frac{9}{4}\gamma$ determine the functional shape of the potential~\cite{soddemann2001generic}, which gives an effective short-range attraction.
Only neutral monomers interact with the additional $V_{\alpha}$ potential, to consider the solvophilic character that charged monomers keep at all temperatures~\cite{del2020charge}.

The equations of motion are integrated through a Nos\'e-Hoover thermostat in the constant NVT ensemble for the equilibration, and through a Velocity-Verlet algorithm in the constant-energy ensemble for the production runs. The integration time-step is $\Delta t = 0.002\tau$, where $\tau = \sqrt{m\sigma^2/\epsilon}$ is the reduced time unit. All simulations are performed with the LAMMPS package~\cite{LAMMPS} at fixed temperature $K_BT/\epsilon = 1.0$.
The particle-particle-particle-mesh method \cite{p3m} has been employed to compute long-range Coulomb interactions. The equilibration of each system is carried out for $2000\tau$, followed by a production run of $6000\tau$ from which we extract the equilibrium averages of the main observables.

\subsubsection*{Main calculated observables}
To assess the swelling properties of the microgels we compute their gyration radius:
\begin{equation}
	R_\text{g} = \sqrt{\sum_{i=1}^{N}\frac{(\vec{r}_\text{i}-\vec{r}_\text{cm})^2}{N}}
\end{equation}
where $\vec{r}_\text{i}$ is the position of the i-th among the $N$ beads and $\vec{r}_\text{cm}$ the position of the centre of mass of the whole microgel.
$R_\text{g}$ gives the overall extension of the mass distribution of the particle and we previously showed that it mainly describes the extension of the core region of the microgel~\cite{del_monte_two-step_2021}. To assess the extension of the full particle, we also calculate the hydrodynamic radius $R_\text{H}$ of the microgel, as~\cite{hubbard1993hydrodynamic}
%To do so, in a recent work~\cite{del_monte_two-step_2021} we showed that this can be calculated from static properties, following Ref.~\cite{hubbard1993hydrodynamic}:
\begin{equation}
\zeta = \frac{K_B T}{\bar{D}} = 6\pi\eta R_H \equiv 6\pi\eta C_\Omega
\label{eq:friction}
\end{equation}
where $\zeta$ is the friction felt by the particle, $\bar{D}$ its diffusion coefficient, $\eta$  the solvent viscosity and $C_\Omega$ an electrostatic capacitance, %\DT{qui l'electrostatic capacitance cade un po' dal cielo per i non specialisti. si puo' dire in una linea che significa?non e' meglio dire "is AN electrostatic capacitance"?}. %which gives a method to estimate the hydrodynamic friction which is linked to  that is determined by its size and 
%This relation is based on a solution of the Navier-Stokes equation for steady flow of rigid particles with stick boundary conditions, where the hydrodynamic interactions are described by the isotropic angular averaged Oseen tensor~\cite{hubbard1993hydrodynamic}.
that is  mathematically equivalent to the hydrodynamic radius, as a good approximation, due to the similarity among the Navier-Stokes equation for the momentum flux density and the Poisson's equation for electrostatics~\cite{hubbard1993hydrodynamic}.
To use this relation in simulations, we approximate the shape of the microgel as an effective ellipsoid~\cite{rovigatti2019elasticity}, to take into account the instantaneous asphericity acquired from the microgel due to the presence of long external chains, and then calculate the hydrodynamic friction for rigid ellipsoids~\cite{hubbard1993hydrodynamic,del_monte_two-step_2021}:
\begin{equation}
R_H=C_\Omega = 2\left[ \int_{0}^{\infty}\frac{1}{\sqrt{(a^2+\theta)(b^2+\theta)(c^2+\theta)}}d\theta \right]^{-1}
\label{eq:capacitance}
\end{equation}
where $a$, $b$, $c$ are the principal semiaxes.

Since we became aware of a different method used in the literature to estimate the hydrodynamic radius of macromolecules, also recently applied to nanogels~\cite{chremos2022influence}, that is based on the ZENO algorithm~\cite{douglas1994hydrodynamic}, we also use this approach to evaluate $R_H$ for all investigated microgels. It is important to note that we find an overall consistency between the two methods, so that both appear to be valid approaches to estimate $R_H$ from implicit solvent simulations.
%In the SI we compare this method with another one, more accurate in the description of the shape of the particle,  finding an overall consistency among the results they give.

To study the internal structure of microgel we also evaluate the radial density profiles, defined as the average densities of a given species $\text{s}$ at a fixed distance from the centre of mass:
\begin{equation}
	\rho_\text{s}(r)= \left\langle \sum_{i=1}^{N_\text{s}} \delta (|\vec{r}_{i}-\vec{r}_\text{cm}|-r) \right\rangle.
\end{equation}
In particular, we monitor the density profiles of all the monomers $\rho_\text{m}(r)$, of charged beads only $\rho_\text{i}(r)$, of the counterions $\rho_\text{c}(r)$ and of the net charge $\rho_\text{Q}(r) \equiv \rho_\text{c}(r) - \rho_\text{i}(r)$.

\section{Results and Discussion}\label{results}
\subsection{Experimental results}
\begin{figure}[ht]
 \centering
    \includegraphics[width=13cm]{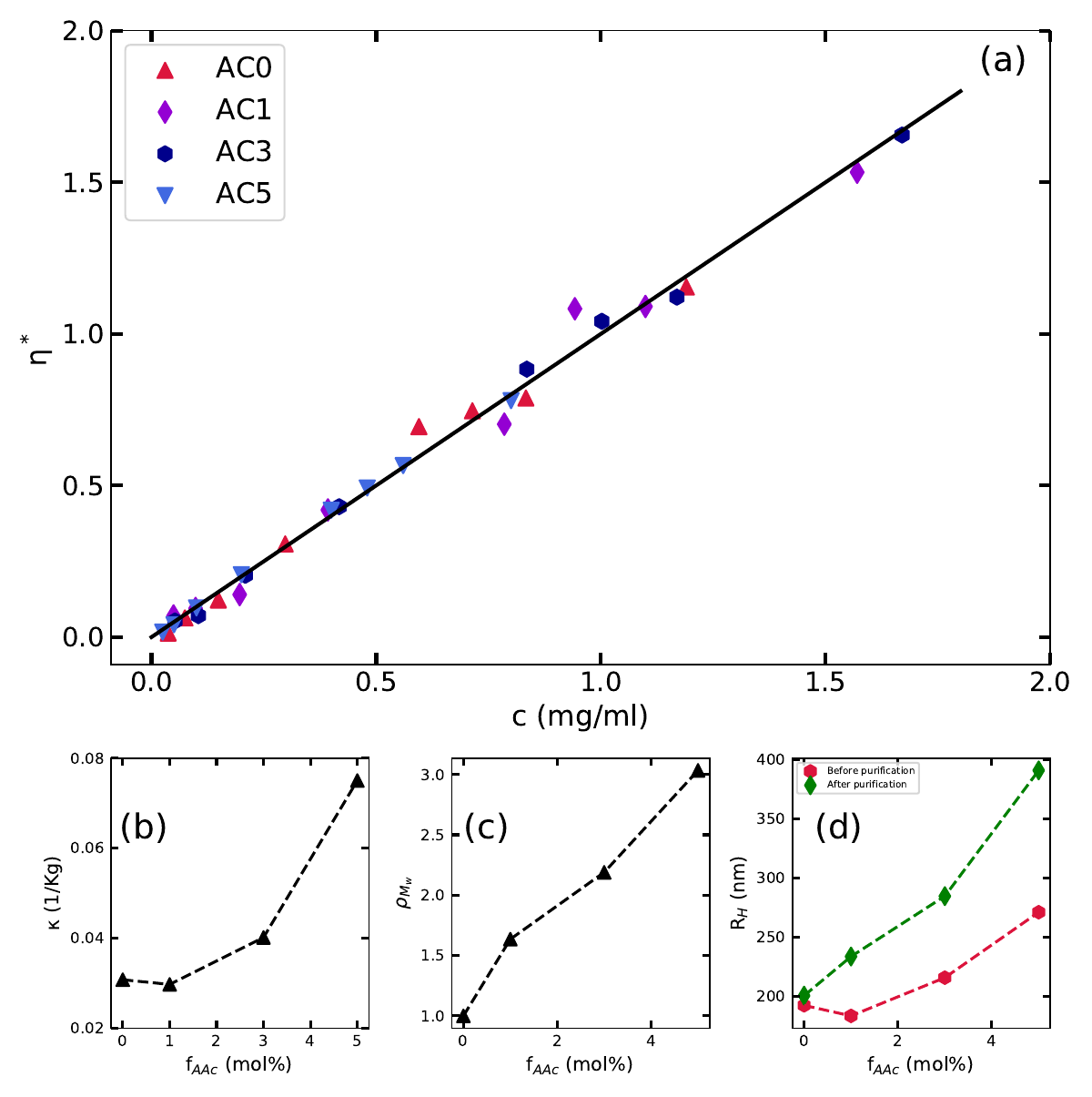}
    \caption{(a) Normalized relative viscosity $\eta^*$ (see Eq.~\ref{eq:etastar}) of all investigated microgels as a function of their mass concentration; (b) Proportionality constant $k$ as a function of AAc fraction; (c) Relative average molecular weight of a single microgel as a function of AAc fraction; (d) Hydrodynamic radius before and after the purification process at $T=20^{\circ}$.}
    \label{fig:Kapa}
\end{figure}

We first discuss the zero-shear viscosity of different microgel suspensions in  the dilute regime since it allows to characterize the role played by AAc monomers during the synthesis in the monomer-to-microgel conversion. As predicted by the Einstein's equation (eq. \ref{eq:Einstein}), $\eta$ increases linearly with particle concentration. All data collapse on the same master curve when the viscosity is appropriately normalized. We thus define a normalized viscosity $\eta^*$ as:
\begin{equation}
   \eta^* = \frac{\eta/\eta_o - 1}{2.5k}.
    \label{eq:etastar}
\end{equation}
Figure \ref{fig:Kapa}(a) reports $\eta^*$ as a function of total polymer concentration for all investigated  samples at 20 $^{\circ}$C, while
Fig.~\ref{fig:Kapa}(b) shows that the proportionality constant $k$ extracted from $\eta(c)$ increases when the fraction of AAc content is augmented in the microgel, similarly to what observed for IPN microgels\cite{franco_volume_2021}, suggesting that for a given number of monomers introduced during the synthesis a higher volume fraction is obtained when the fraction of acrylic acid units increases. 
 At this stage we cannot discern whether this is due to a higher number of microgels or to a larger microgel mass. To overcome such difficulty we can couple the so obtained $k$ values to the hydrodynamic size measured via DLS at the same temperature as in viscosimetry tests (20 $^{\circ}$C, Table \ref{tbl:Summary}) and measure the relative variation of their molecular weight with respect to a reference sample. In more detail, since $k=\varphi/c$, $\varphi=N_mV_m/V$ and $c=M/V$, where $N_m$ is the total number of suspended microgels, $V_m=(4/3 \pi R_H^3)$ the volume of one microgel, $V$ the volume of the suspension and $M$ the total mass of monomers introduced during the synthesis, we straightforwardly get
\begin{equation}
   \frac{V_m}{k}=\frac{M}{N_m}=M_w^E
    \label{mw}
\end{equation}
where the ratio $\frac{M}{N_m}$ is nothing but an effective molecular weight of a microgel, whose value cannot be considered strictly correct, since it leverages on equation \ref{eq:Einstein}, where the factor 2.5 holds for hard spheres, while here we have penetrable and deformable colloids. Nonetheless we can compare results obtained for $M_w^E$ with respect one reference sample, AC0 here, and define the molecular weight ratio $\rho_{M_w}=\frac{M_w^E(f_{AAc})}{M_w^E(f_{AAc}=0)}$, that is not \emph{a priori} dependent on the specific first order correction to the viscosity of the solvent due to the presence of penetrable microgels for a given microgel structure.
Moreover, despite a softening due to less compact structures of the microgels and a reduction of the hydrodynamic friction coefficient for increasing $f_{AAc}$ is still possible, this would give rise to a decrease of $k$, that is not emerging from our analysis.

The molecular weight ratio $\rho_{M_w}$ is reported in figure \ref{fig:Kapa}(c) as a function of the acrylic acid fraction. We find that $\rho_{M_w}$ increases as a function of the AAc content in the microgels, implying that the mass of a single particle increases for increasing $f_{AAc}$. This occurs despite the total number of monomers is kept constant in all the syntheses (see SI). Such an increase in the microgel mass can be explained only if i) the number of microgels and nucleation sites decreases or ii) the number of unreacted monomers progressively decreases with increasing $f_{AAc}$. We rule out the second hypothesis since we do not detect any significant variation of the residual mass obtained by drying an aliquot of supernatant resulting after one centrifugation cycle: the unreacted mass was always about 1\% of the total NIPAm dissolved in water during the synthesis. Therefore we do not expect an impact of monomer conversion on the final mass of a single microgel. This suggests that the observed behavior is due to a decrease in the number of microgels. These findings are consistent with the fact that, at the synthesis temperature (70 $^{\circ}$C), the propagation rate of NIPAm monomers is much higher than that of acrylate units \cite{kroger_prediction_2017, beuermann_rate_2002}, the latter being  more hydrophilic and partially charged at all temperatures in water. We thus speculate that this difference makes AAc monomers less reactive units during the synthesis, decreasing in turn the number of nucleation sites for the chain growth and hence increasing the final molecular weight of the crosslinked networks.
The larger molecular weight and hydrophilicity of highly charged microgels is further corroborated by the comparison between the hydrodynamic radii measured before and after purification (Figure \ref{fig:Kapa}-(d)). The difference between the two sizes progressively increases for increasing $f_{AAC}$, showing that purification induces a larger swelling in microgels containing more charged groups and characterized by a larger molecular weight. We ascribe such an enhanced swelling mainly to a net reduction of the electrostatic screening and AAc protonation due to the removal of leftover SDS, unreacted AAc and NIPAm monomers and part of their respective free counterions.           

%The increase in $k$ with increases in AAc content implies that the mass of a single microgel in the suspension increases with increase in AAc content. These results initially sound contradictory since the molar mass of AAc is lower than that of PNIPAm and therefore a increase in AAc content should decrease the total mass of a single microgel in the suspension. 

\begin{figure}[ht]
 \centering
    \includegraphics[width=17cm]{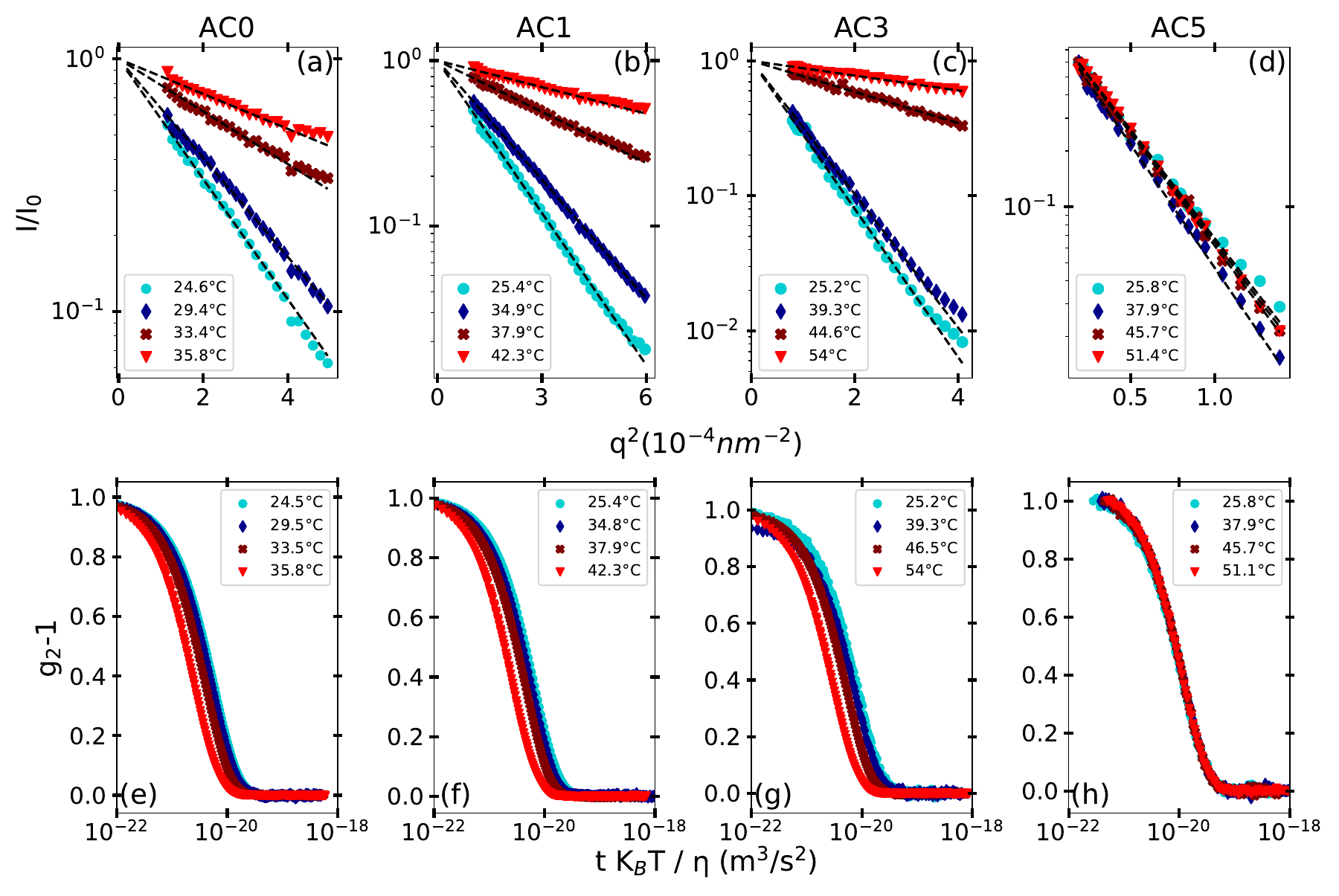}
    \caption{Normalized scattered intensity $I(q)$ from static light scattering (panels, a,b,c,d) and autocorrelation functions $g_2-1$ from dynamic light scattering (panels e,f,g,h) for all studied microgels (AC0, AC1, AC3, AC5) at different temperatures.}
    \label{fig:SLSDLS}
\end{figure}

\begin{table}

  \caption{Summary of the results from static and dynamic light scattering for all the microgel samples: Hydrodynamic radii at 20 $^{\circ}$C, critical temperatures $(T_c^H,T_c^g)$ and sharpness parameters $(s^H,s^g)$ obtained by fitting $R_H(T)$ and $R_g(T)$ via equation \ref{eq:critical fit}.}
  \label{tbl:Summary}
  \begin{tabular*}{\textwidth}{@{\extracolsep{\fill}}lllllll}
    \hline
    sample code &$f_{AAc}$ & $R^{20}_H$ (nm) & $T_c^{H}$ ($^{\circ}$C) & $T_c^{g}$ ($^{\circ}$C) & s$^H$ ($^{\circ}C^{-1}$) & s$^g$ ($^{\circ}C^{-1}$)\\
    \hline
    AC0 & 0\% & 200.5 & $33.70 \pm 0.03$ & $33.06 \pm 0.07$ & $0.71 \pm 0.03$ & $1.39 \pm 0.30$ \\
    AC1 & 1\% & 233.6 & $38.62 \pm 0.06$ & $36.94 \pm 0.04$ & $0.32 \pm 0.01$ &$0.50 \pm 0.02$ \\
    AC3 & 3\% & 284.6 & $46.81 \pm 0.14$ & $43.52 \pm 0.05$ & $0.24 \pm 0.01$ &$0.49 \pm 0.02$ \\
    AC5 & 5\% & 390.8 & -                &  -               &   -             &- \\
    \hline
  \end{tabular*}
\end{table}

\begin{table}

  \caption{Summary of results from mobility measurements: mobility $\mu$ and normalized mobility $\mu\eta_0/\epsilon$ measured at 20 $^{\circ}$C and 60 $^{\circ}$C for all the investigated samples. The reported errors are the standard deviations obtained over multiple runs.}
  \label{tbl:Summary1}
  \begin{tabular*}{\textwidth}{@{\extracolsep{\fill}}lllll}
    \hline
    \small sample code & \footnotesize$\mu^{20}$ \footnotesize($10^{-8}m/Vs$) & \footnotesize$\mu^{60}$ \footnotesize($10^{-8}m/Vs$) & \footnotesize$\mu^{20}\eta_0/\epsilon$ \footnotesize($10^{-8}m$ $Pa/Vs$) &\footnotesize$\mu^{60}\eta_0/\epsilon$ \footnotesize($10^{-8}m$ $Pa/Vs$) \\
    \hline
    AC0& -0.87 $\pm$ 0.04 & -5.90 $\pm$ 0.29 & -0.010 $\pm$ 0.001 & -0.041 $\pm$ 0.002\\
    AC1& -0.58 $\pm$ 0.03 & -6.05 $\pm$ 0.30 & -0.007 $\pm$ 0.001 & -0.042 $\pm$ 0.002\\
    AC3& -1.16 $\pm$ 0.06 & -5.85 $\pm$ 0.29 & -0.014 $\pm$ 0.001 & -0.041 $\pm$ 0.002\\
    AC5& -2.40 $\pm$ 0.12 & -5.13 $\pm$ 0.26 & -0.030 $\pm$ 0.002 & -0.036 $\pm$ 0.002\\
    \hline
  \end{tabular*}
\end{table}

Figure \ref{fig:SLSDLS} shows the results obtained from both dynamic and static light scattering experiments. The top panels (a,b,c,d) report $I(q)/I_0$ as a function of $q^2$ for selected temperatures and all  investigated AAc fractions. 
In all cases we observe an exponential decay in a quite large window of scattering vectors allowing to fit the intensity via equation \ref{eq:guiner} and extract $R_g$. The presence of the microgel VPT, signalled by a large variation of the slope of $log(I/I_0)$, is evident for all the microgels but one, namely the  AC5 sample, where $I(q)/I_0$ does not appreciably change with temperature. We also note that $I(q)/I_0$ decays more rapidly for increasing wave vectors $q$ as the AAc content of microgels grows, pointing to a net increase of the gyration radius of the microgels embedding more acrylic acid units. This occurs essentially at low temperatures where all microgels are swollen.    
%\DT{Already said}.$R_g$, which describes the standard deviation of the mass distributed around the centre of mass, was extracted using the Guiner approximation (equation \ref{eq:guiner}). 
%The slope of $I(q)/I_0$ vs $q^2$ with temperature increases algebraically as the AAc content of microgels increases. When $f_{AAc} = 5\%$ is reached, $I(q)/I_o$ does not change with temperature implying that the $R_g$ does not undergo a collapse at high temperature.  

The autocorrelation functions ($g_2(t)-1$) obtained via dynamic light scattering for samples with different AAc content are reported in figure \ref{fig:SLSDLS} (bottom panels, e,f,g,h). The lag time $t$ is normalized by $K_BT/\eta$ to filter out the trivial speeding up of the dynamics due to the temperature dependence of the solvent viscosity and to the thermal agitation. Hence, the obtained shift at different temperatures is only due to the change in $R_H$ of the microgels. Once again, in agreement with the SLS data discussed above, for $f_{AAc}=5\%$ the intensity autocorrelation functions do not show any remarkable variation with temperature, indicating that $R_H$ of the microgels does not appreciably change in the accessible range of temperatures.   

\begin{figure}[ht!]
 \centering
    \includegraphics[width=13cm]{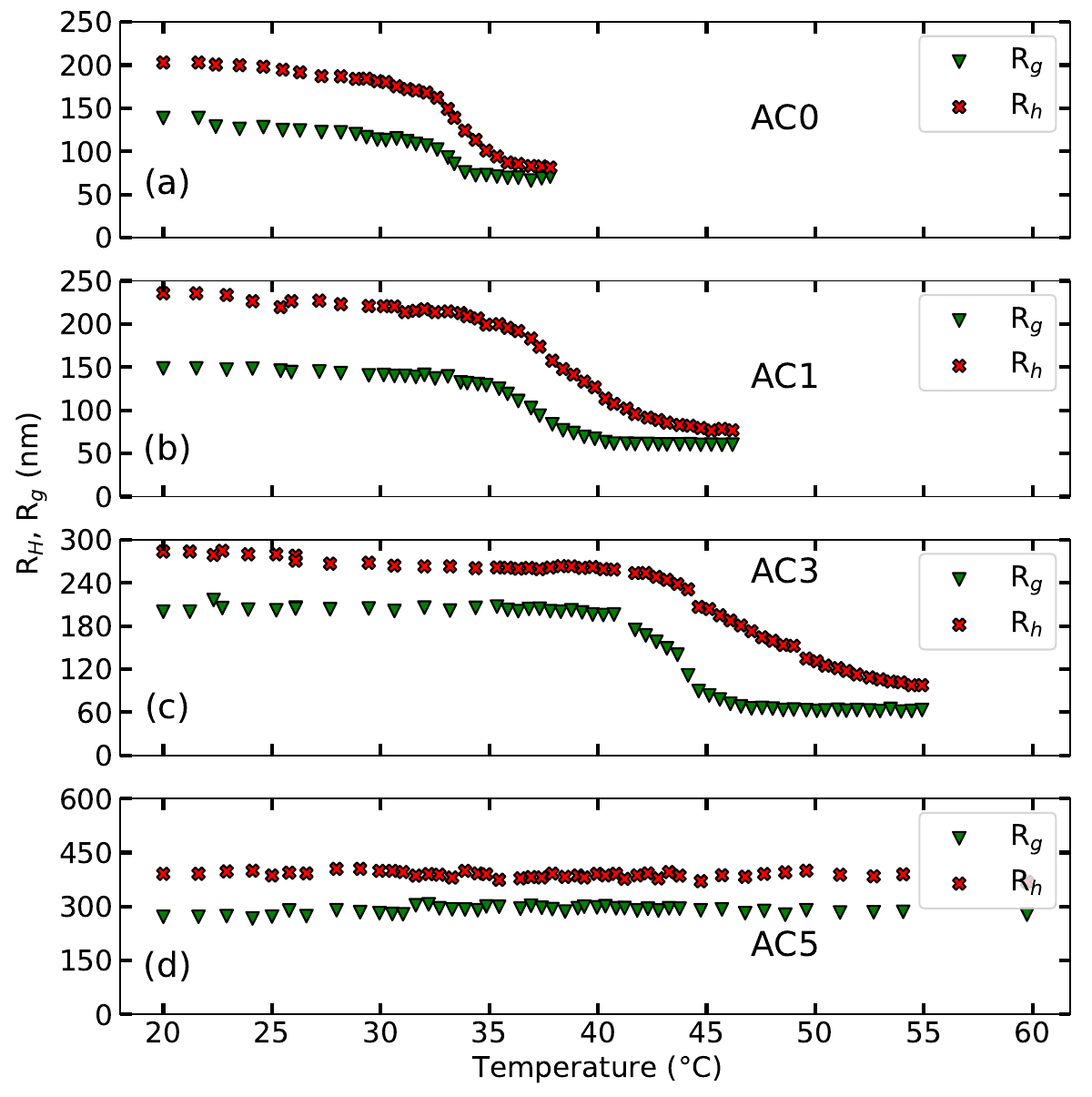}
    \caption{Hydrodynamic radii R$_H$ and gyration radii R$_g$ of PNIPAm-co-AAc microgels as a function of temperature for varying concentration of acrylic acid (f$_{AAc}$). (a) AC0, (b) AC1, (c) AC3 and (d) AC5}
    \label{fig:RgRh}
\end{figure}

The resulting $R_g(T)$ and $R_H(T)$ for all investigated microgels are shown in figure \ref{fig:RgRh}. Both quantities increase below $T_c$ for increasing AAc content, showing that acrylic acid addition enhances the overall hydrophilicity of the microgels and it is consistent with an increase of their mass as well. We may in principle also hypothesize that, since the synthesis of the microgels occurs at $70$ $^{\circ}$C, i.e. a temperature at which the PNIPAm monomers become partially hydrophobic, and a fraction, even very small, of AAc monomers always stay  deprotonated also at low pH, more AAc units could tend to occupy the periphery of the microgel (see also discussion below). However we anticipate here that this does not result in a detectable change of the swollen microgel structure as witnessed by the nearly constant ratio of the two measured sizes ($R_g/R_H\simeq$ 0.7) at 25 $^{\circ} C$ for all acrylic acid fraction. The thermal evolution of such a ratio will be discussed afterwards. A strongly uneven distribution of charged acidic groups therefore has to be excluded, suggesting that the already mentioned increase of $R_H$, once the synthesis leftovers are removed by centrifugation/supernatant removal cycles (figure \ref{fig:Kapa}-d), must be attributed to the larger mass and the augmented electrostatic self-energy of the entire AAc-rich microgels, and not only to the swelling of the peripheral corona.  

As the temperature is increased, microgels undergo a volume phase transition due to the presence of a finite Low Critical Solution Temperature (LCST) for PNIPAm. However, the electrostatic repulsion among the AAc-rich chains, together with their hydrophilic character, tends to keep the microgel swollen, thus shifting the occurrence of the VPT to higher $T$-values. The critical temperatures ($T_c$) at which the microgels undergo the transition are estimated by fitting the both $R_g$ and $R_H$ to an auxiliary function \cite{del_monte_two-step_2021}:
\begin{equation}
   R_{H,g} = [R_0 - \Delta R_{H,g} \tanh(s(T-T_c^{H,g}))]+ A(T-T_c^{H,g})
    \label{eq:critical fit}
\end{equation}
where, $R_0$ is the radius of the microgel at the VPT, $\Delta R_{H,g}$ is the amplitude of the VPT and the parameter $s$ quantifies its sharpness. The obtained values of $T_c$ for the different samples are reported in Table \ref{tbl:Summary} for both $R_g$ and $R_H$, while the best-fitting curves are shown in SI. As foreshadowed by Fig.~\ref{fig:SLSDLS}, $T_c$ shifts to higher temperature as the AAc content increases, until for $f_{AAc}=5\%$ neither $R_g$ nor $R_H$ show any sign of collapse: it appears that the gain in translational entropy of water that would result in the collapse of these microgels is not sufficient anymore to overcome electrostatic repulsion between AAc-rich chains. Such a scenario is further supported by the electrophoretic mobility data $20$ $^{\circ}$C and $60$ $^{\circ}$C, reported in Table~\ref{tbl:Summary1}, which confirm the presence of AAc monomers and charged initiators within the microgels and are sensitive to microgel collapse. We observe a net increase (in modulus) of the absolute mobility $\mu$ for the AC3 and AC5 samples when compared to AC0, while we ascribe the weak decrease of the mobility in AC1 sample to the enhanced screening due to the presence of confined AAc counterions. Whether the electrophoretic mobility of swollen PNIPAm-co-AAc microgels shows a non-monotonic behavior in function of the acrylic acid content due to the subtle balance between electric and frictional forces of different origin\cite{obrien_electrophoretic_1978,zhou_computer_2015} is currently unknown, although a non-trivial behavior of the mobility has already been reported as a function of salt concentration\cite{truzzolillo_overcharging_2018}. This represents an interesting aspect \emph{per se} and will not be investigated further here.     
In all cases we observe a large increase of mobility at high temperature. To discern whether this is due to their collapse or simply to the relative variation of the viscosity $\eta_0$ and/or the relative permittivity $\epsilon$ of the solvent\cite{hunter2013zeta}, we also report the normalized mobility $\mu\eta_0/\epsilon$ for the two temperatures. We unambiguously observe that the normalized mobility for AC0, AC1 and AC3 increases by factor whose value is comprised between 3 and 4, going from 20 $^{\circ}$C to 60 $^{\circ}$C, while for the AC5 sample it stays almost unaltered, corroborating the absence of significant deswelling for this sample. 
Another important feature due to the addition of AAc is the decrease of the "sharpness" parameter $s$ for increasing acrylic acid fraction $f_{AAc}$ (Table \ref{tbl:Summary}). This can be attributed to the increase of repulsive units into the temperature-sensitive NIPAm structure, pointing to a relatively even distribution of acidic "defects" within the microgels, which reflect a net decrease of the cooperativity of the volume phase transition: the latter is characterized by a larger spectrum of critical temperatures for increasing AAc content. It is also worth noting that, despite we observe a clear broadening of the volume phase transition, we do not observe a successive sequence of transitions as those described in previous works\cite{kratz_influence_2000,huo_temperatureph_2006} for larger $f_{AAc}$, which reported first a collapse of PNIPAm-rich domains, followed by the collapse of the AAc-rich domains at higher temperatures. Such a scenario has been previously rationalized by proton migration from the solution to the microgels after the first collapse \cite{kratz_influence_2000}, inducing a large counterion condensation on the carboxylic groups and a second microgel deswelling.  Our results therefore suggest a homogeneous distribution of acrylic acid segments and continuous transitions, as discussed in \cite{karanastasis_one-pot_2020} for PNIPAm-co-Methacrylic acid microgels.  Transition broadening and VPT suppression have been observed for increasing fraction of deprotonated acrylic acid groups also in very similar syntheses of PNIPAm-co-AAc microgels \cite{al-manasir_effects_2009,farooqi_stability_2017}.
In addition, since during the synthesis (before purification) the nominal pH of the samples is less than pKa$^{(AAc)}$ (see section \ref{matmeth}), their high protonation degree could
%\gdm{Is this pH resulting only from the dissociation of AAc or was the synthesis conducted in acidic conditions? Among the introduction and methods I found two contradictory sentences, where we say that AAc is protonated, in one case, and deprotonated, in the other. We should harmonize the text, but I can't remember the whole story about pH, among synthesis and measurements\ldots}\DT{I should have done it. See also Methods/Synthesis} 
increase their solubility into the NIPAm droplets formed at 70 $^\circ$C and homogenize their distribution within the emerging network. This said, we cannot exclude that two distinct collapses of PNIPAm-rich and AAc-rich domains might occur at higher pH ($>$6.9), namely for larger fractions of deprotonated AAc monomers. This hypothesis will be a subject of future investigation.
%\gdm{Isn't all the AAc always dissociated?} \gdm{I would like to discuss the point of two step collapse in the reference~\cite{kratz_influence_2000}, because prof. Hellweg, if I remember well, during a talk once told me that in those microgels AAc should be considered as randomly spread (but I have to give again a look to the paper, to be sure\ldots)}
%\ez{per me la discussione puo andare  e la chiuderei qui...}

%It is also important to point out that the VPT shown in figure \ref{fig:RgRh} is only due to the change in temperature and not to the change in pH (which is also another parameter that can lead to a VPT \Rajam{ref to be added}) since the pH of the samples was not modified during the temperature ramps. \DT{not very clear}.

\begin{figure}[ht]
 \centering
    \includegraphics[width=13cm]{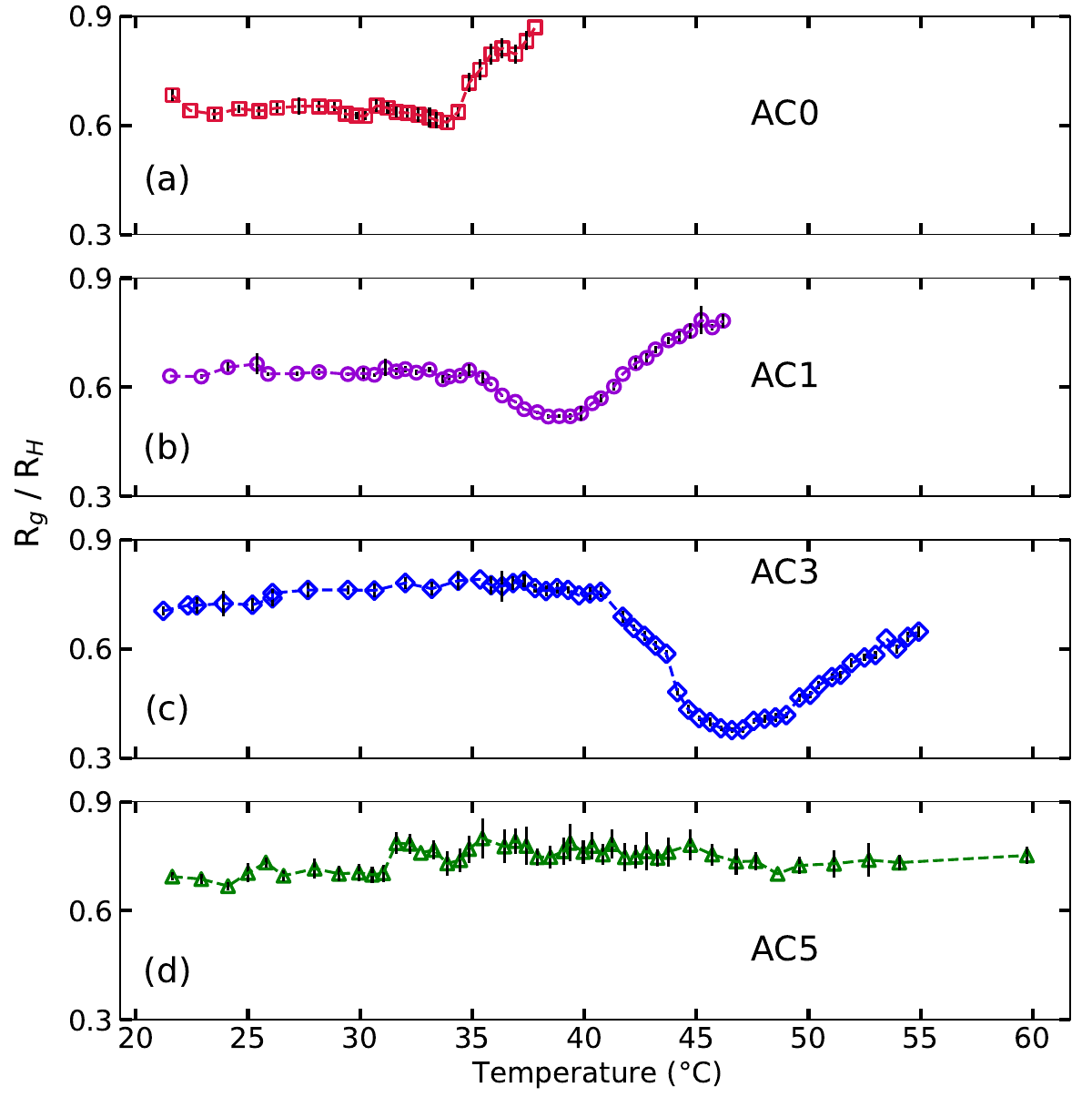}
    \caption{Ratio of R$_g$/R$_H$ of PNIPAm-co-AAc microgels as a function of temperature for all the investigated samples as indicated in the panels.} 
    \label{fig:Ratio}
\end{figure}

As clearly visible in Fig.~\ref{fig:RgRh}, $R_g$ is systematically lower than $R_H$ for all samples and the decrease of both quantities with temperature is more and more decoupled upon increasing the fraction $f_{AAc}$. The decoupling in the VPT for $R_H$ and $R_g$ can be  clearly seen by looking at the difference between the critical temperatures ($\Delta T_c = T^H_C - T^g_C$) obtained for the two quantities. For AC0 microgels, $\Delta T_c$ is $\le 0.7^{\circ}C$, and it increases up to $\Delta T_c\approx 3.3$ $^\circ$C for AC3 (see Table \ref{tbl:Summary}). Therefore, for increasing $f_{AAc}$, the transitions of $R_H$ and $R_g$ become more and more distant in temperature. Such an enhanced decoupling results in the presence of a minimum in the ratio $R_g/R_H$ as a function of temperature, as shown in Figure~\ref{fig:Ratio}.   
At low temperatures, all samples exhibit a ratio $0.6\leq Rg/R_H\leq 0.8$ until $T_C$ is reached. Since the transition in $R_g$ anticipates that in $R_H$, a minimum appears when $R_g$ collapses, with a minimum depth depending on $f_{AAC}$, coherently with the $\Delta T_c$ trend. The added charges thus enhance the inhomogeneous deswelling and the two-step character of the volume phase transition of thermosensitive microgels, a feature that up to now has been identified only in pure PNIPAm microgel suspensions~\cite{del_monte_two-step_2021}. Up to our knowledge this is the first experimental evidence of the influence of charged AAc co-monomers on the structural change across the continuous VPT of PNIPAm-based microgels.  

\subsection{Simulations}
To provide a better understanding of the role played by the presence of co-polymerized ionizable groups in the VPT of microgels, we performed molecular dynamics simulations as described in section \ref{subsec:simulations}.                               
\begin{figure}[ht]
 \centering
    \includegraphics[width=\textwidth]{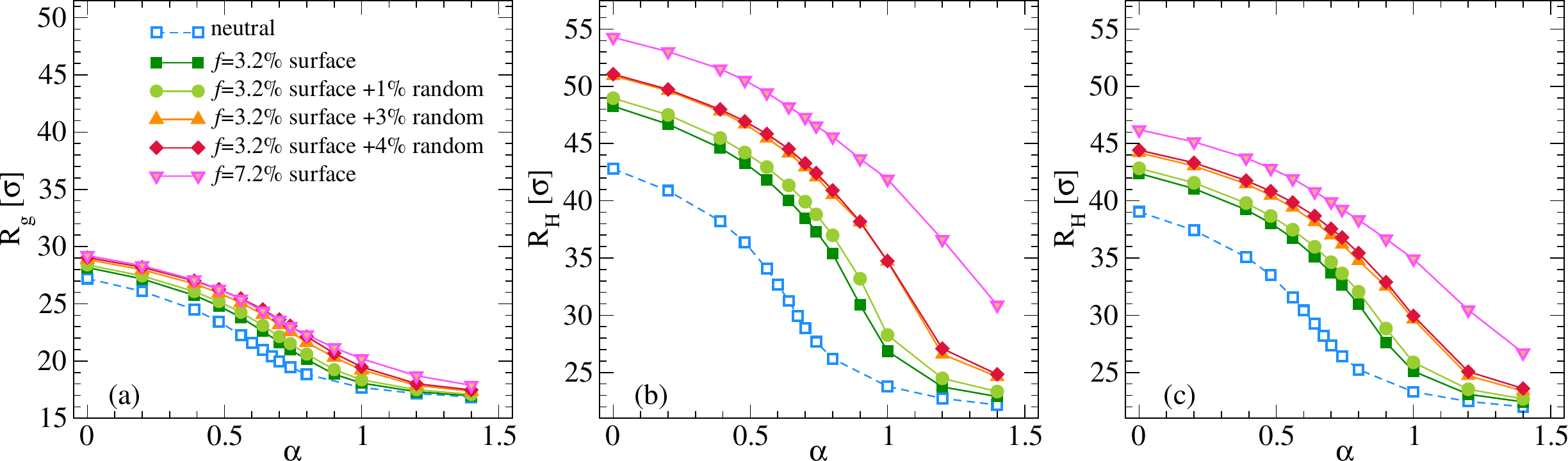} \\
    \vspace{0.5cm}
    \includegraphics[width=\textwidth]{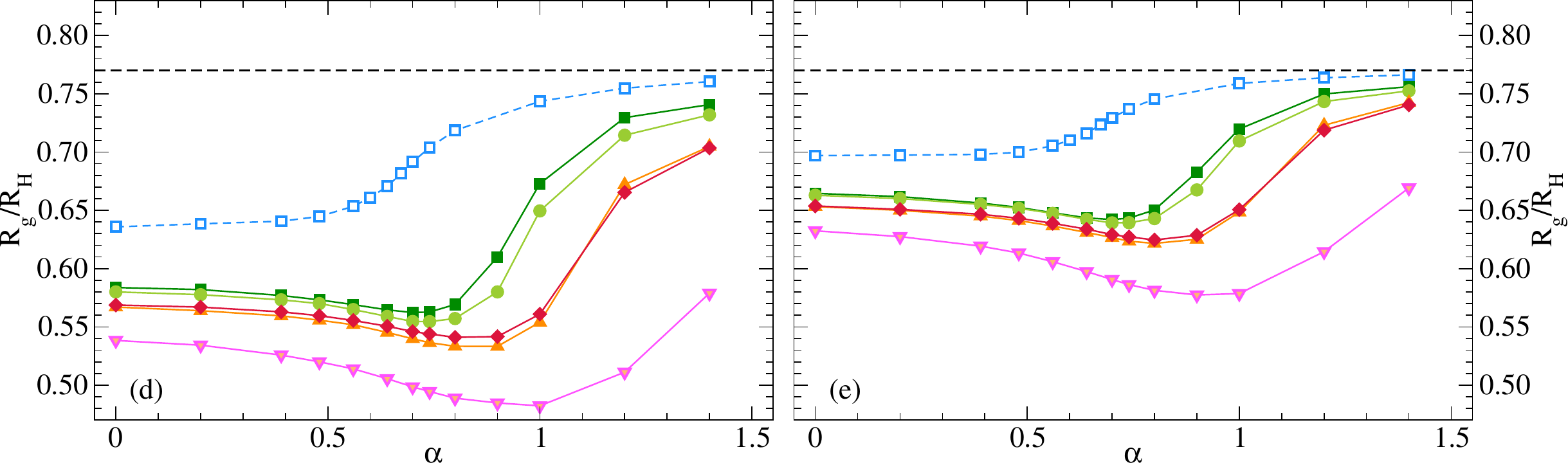}
    \caption{Top: Swelling curves of (a) $R_g$, (b) $R_H$ computed with the convex hull method and (c) $R_H$ computed with the ZENO algorithm.
     Bottom: Swelling ratios among $R_g$ and $R_H$, as computed with the (d) convex hull method and (e) the ZENO algorithm.
     }
    \label{fig:sim:swelling}
\end{figure}
Fig.~\ref{fig:sim:swelling} shows the results for (top row) the radius of gyration (a) and for the hydrodynamic radius, calculated through the convex hull method (b) and the ZENO algorithm (c), respectively, 
In agreement with experiments, for all charge contents and distributions, the gyration radius of the microgels is smaller than the hydrodynamic radius at all the temperatures, and the increase of charge results in a net increment of the microgel size, with the hydrodynamic radius being more affected than the gyration one. Most importantly, the hydrodynamic volume phase transition temperature experiences a larger shift as the fraction of ionized beads is increased with respect to the same transition temperature extracted from the gyration radius data: the two transitions decouple because of the charge addition. We note that the results obtained for $R_H$ from the two methods are in full qualitative agreement, giving us confidence in both of them for a reliable estimate of the hydrodynamic radius of macromolecules in solvent-free molecular dynamics simulations. 

Such an agreement is also observed in the behavior of the ratio $R_g/R_H$ ratio, plotted as a function of the solvophobicity parameter $\alpha$ in Fig.~\ref{fig:sim:swelling}(d,e), respectively for the convex hull and the Zeno methods. 
The observed decoupling in the transitions of $R_g$ and $R_H$ is thus reflected in the appearance and in the successive deepening of a minimum in their ratio. We remark that this feature is strictly absent for neutral microgels for both methods, stressing once more the importance of an appropriate charge modeling to reproduce the experimental behavior even in pure PNIPAm homopolymer microgels. 
At this point, it is worth noticing that in experiments it is hard to obtain strictly neutral PNIPAm-based microgels in aqueous environments, since non-ionic initiators are i) poorly soluble in water, making the synthesis problematic and hardly reproducible, and ii) often show successive unwanted deprotonation, as in the case of 2-2'-azobisisobutirronitrile (AIBN) that produces anionic microgels as  already discussed  in other microgel synthesis procedures\cite{bonham_role_2017}. In this respect, simulations appear to be crucial since they put forward the evidence that the two-step deswelling is strictly an electrostatic phenomenon.  
In addition, the minimum is further affected by the spatial charge distribution. Despite the presence of only 3.2\% of charged beads within the outermost corona of the microgels is enough to trigger the aforementioned decoupling and the subsequent minimum in $R_g/R_H$, the increase of the amount of ionic beads %randomly distributed within the microgel 
enhances such a decoupling mechanism and deepens the minimum. This effect is already present when additional charged groups are randomly distributed within the microgel, but becomes much more evident when charges are all distributed on the surface.
 A tentative explanation of this effect, when adding charges randomly throughout the network, would be that those groups located in the core are always well screened by counterions, thus barely affecting the $R_g$ transition, while the fraction of them that still goes to the surface results in an increase of the charged layer of the corona, as Figure \ref{fig:sim:structure:a0.80} shows, in turn increasing its swelling and the transition temperature of $R_H$. The more charged the surface, the more inhomogeneous is the structural swelling transition.
This is confirmed by the results obtained with 7.2\% of ionic beads all distributed within the corona: in this case,  the decoupling between $R_g$ and $R_H$ is enhanced, allowing us to conclude that charge accumulation in the corona is the main driving mechanism for the observation of the two-step deswelling, as also shown by the large increase of the minimum depth for this particular case (Figure \ref{fig:sim:swelling}-d,e).
It is important to note that this feature is also accompanied by a shift of the minimum position to larger solvophobic parameter values, hence effective temperature, in good agreement with experiments (Figure \ref{fig:RgRh}) for increasing amount of acrylic acid monomers.
%However, since we suspect that also the acrylic acid monomers may be preferentially located in the microgel periphery we performed a further set of simulations to compare two configurations having the same amount of charges with different spatial distribution. For this purpose we simulated a microgel with 7.2\% of ionic beads distributed uniquely in the outer corona that has therefore the same fraction of charges $f$ as the microgel with 3.2\% of ionic beads on the surface and 4\% of ionic beads randomly distributed within the microgel volume. 
In simulations we also observe a mild decrease of the ratio $R_g/R_H$ at low effective temperatures (swollen microgels) between the two configurations just mentioned and more generally with increasing $f$. This is not observed in experiments and could be due to the fact that our charged groups are added to the same microgel topology, assembled {\it in silico} in the absence of charges, which may result in a slightly different structure of the simulated networks in the swollen state. However, this aspect goes beyond the scope of the present work and will not be investigated further.
%we impute that to ....\DT{vogliamo commentarlo?any idea? a me viene da dire: we impute that to the microgel synthesised in silico characterized by a more compact core with respect to that of microgels used in experiments in the swollen state. This aspect however goes beyond the scope of the present work and will not be investigated further} \gdm{yes, I also think that is for a similar reason... in the sense that, maybe, when we put a lot of charged beads we make a strong perturbation in the core-corona structure obtained through our in-silico assembly. Probably the same real structure of microgels is different, when a lot of AAc is added, hence it should be accounted in the assembly procedure\ldots BTW, I found very very difficult to find in literature complete form factors of co-pol mgels, and once prof. Hellweg said that is because of their typical size, but it would be difficult to make the same work done with pNIPAM microgels with Jerome, I think\ldots}. 
Our simulations thus confirm a general scenario where electrostatic charges in thermosensitive microgels give rise to a two-step deswelling process with the core collapsing first than the outer corona and the ratio $R_g/R_H$ starting from values comprised in the interval 0.6-0.7 at low temperatures, passing through a minimum and finally tending to the hard sphere limit ($R_g^{HS}/R_H^{HS}=0.77$) when the collapsed state is approached.
\begin{figure}[h!]
 \centering
    \includegraphics[width=0.47\textwidth]{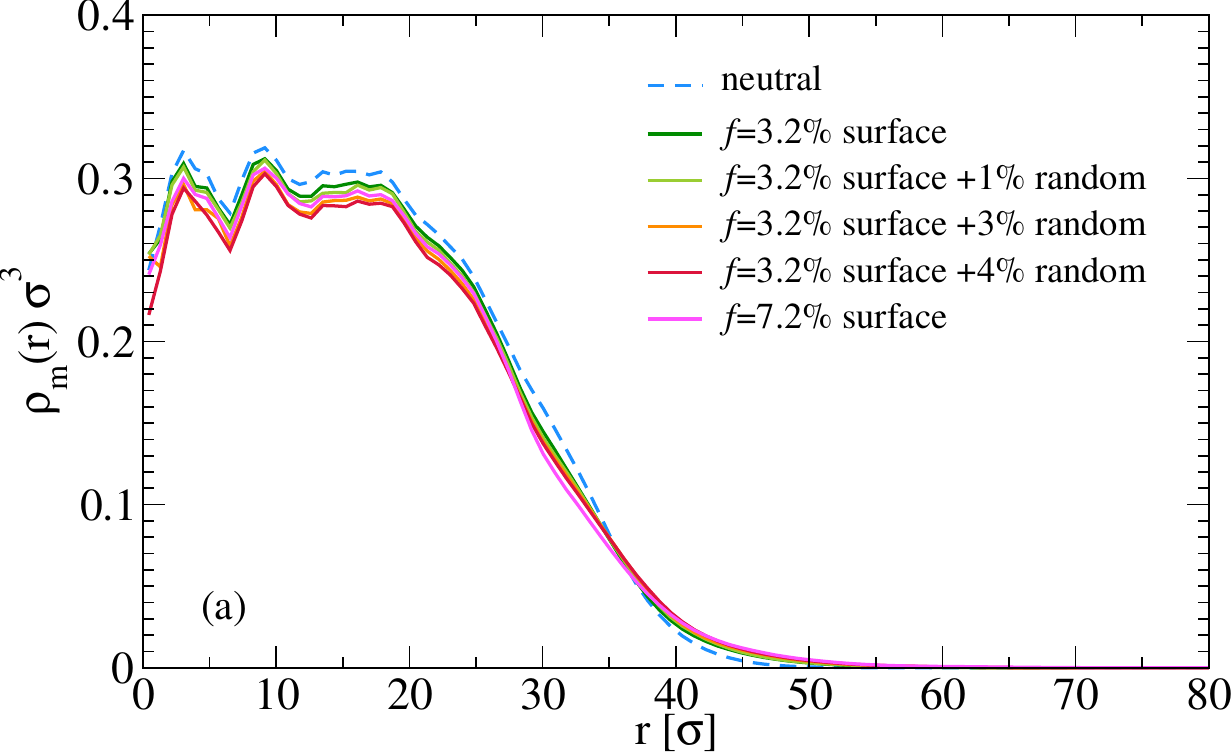}
    \hfill
    \includegraphics[width=0.47\textwidth]{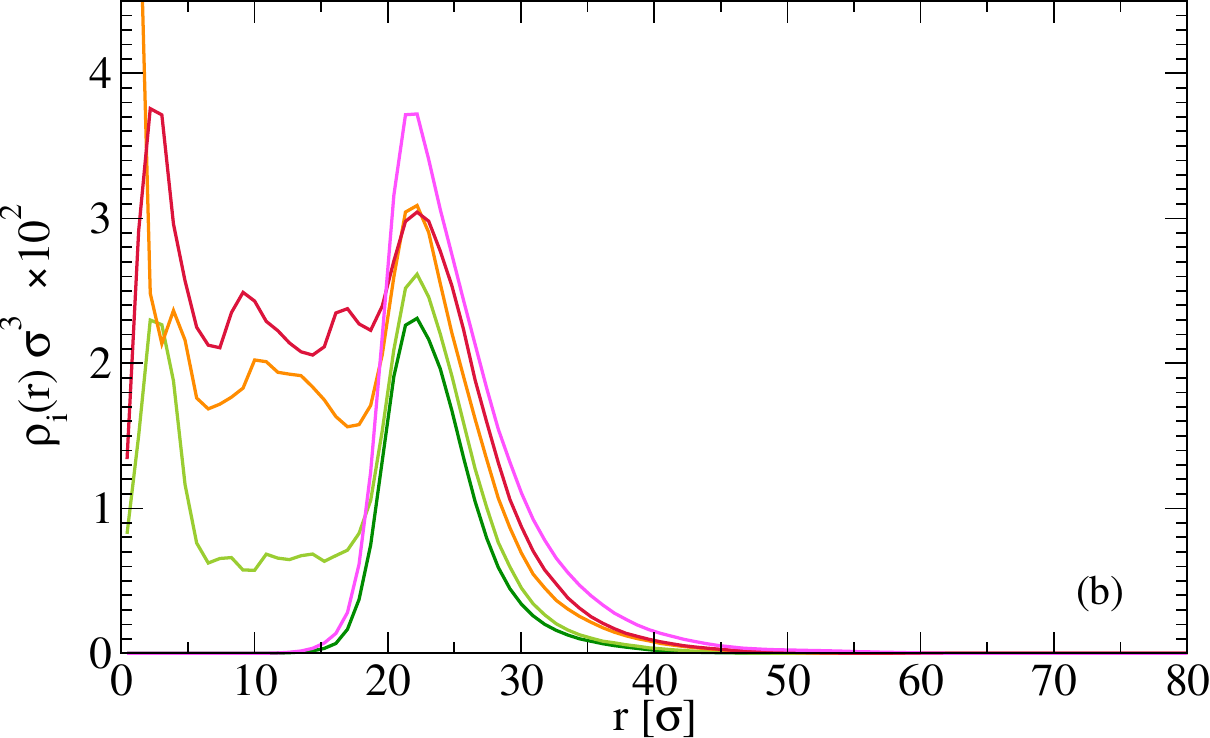}\\
    \vspace{0.5cm}
    \includegraphics[width=0.47\textwidth]{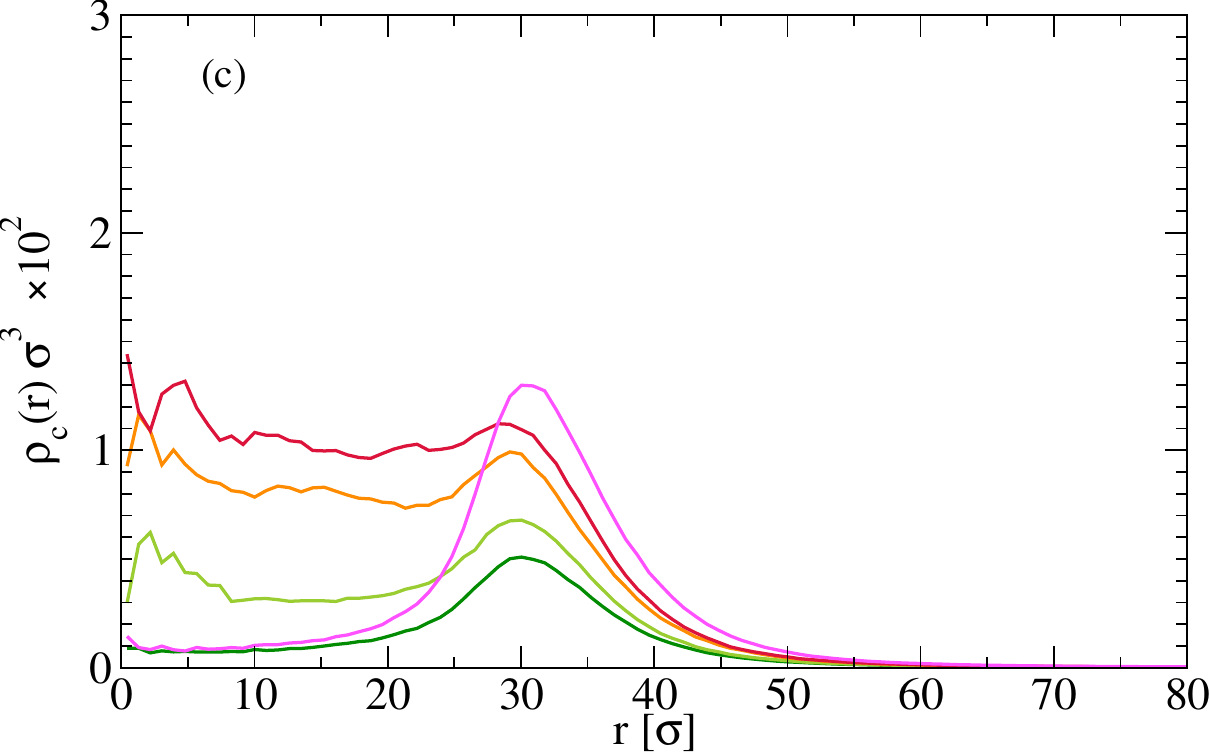}
    \hfill
    \includegraphics[width=0.47\textwidth]{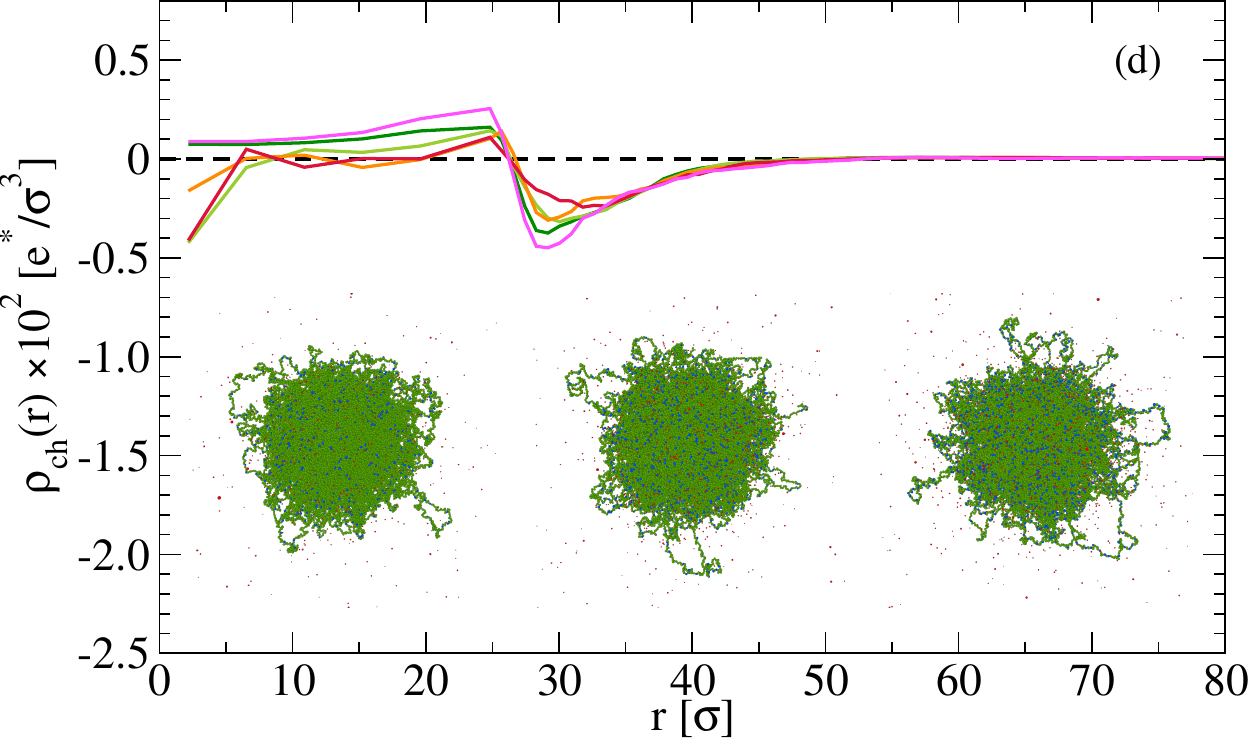}
    \caption{(a) Monomers, (b) ions, (c) counterions and (d) net charge radial density profiles for the microgels as a function of the distance from the microgel center of mass $r$ for $\alpha=0$. Inset in (d): snapshots corresponding to the microgels with surface charge $f_s=3.2\%$, with surface charge $f_s=3.2\%$ and random charge $f_r=4\%$, and with surface charge $f=7.2\%$, from left to right.}
    \label{fig:sim:structure:a0.00}
\end{figure}

In order to have a closer look to the effect of the VPT on the charge distribution (and viceversa), we report in Figs.~\ref{fig:sim:structure:a0.00}, ~\ref{fig:sim:structure:a0.80} and ~\ref{fig:sim:structure:a1.20} the density profiles of all monomers (a), charged ions (b), counterions (c) and net charge (d) at three representative temperatures, $\alpha=0, 0.80$ and $1.20$ corresponding to the fully swollen, nearly critical and collapsed state of the microgels, respectively. %The profiles are also supported by corresponding microgel snapshots for few significant charge distributions.

For swollen microgels ($\alpha=0$) the monomer density (panels (a)) is only weakly affected by ion embedding. We observe that the tail of the distribution become slightly fatter for increasing $f$ or net charge mismatch between the corona and the core, while its $(r\rightarrow 0)$ value of the density profile decreases. Such a weak, albeit detectable effect, can be attributed to the augmented microgel size. This feature also holds for the other two values of $\alpha$ shown in Figs.~\ref{fig:sim:structure:a0.80},\ref{fig:sim:structure:a1.20}(a), being more evident for near-critical microgels ($\alpha=0.8$) where the effect of the added charge is magnified, along with inhomogeneities in the structure.
\begin{figure}[ht]
 \centering
    \includegraphics[width=0.47\textwidth]{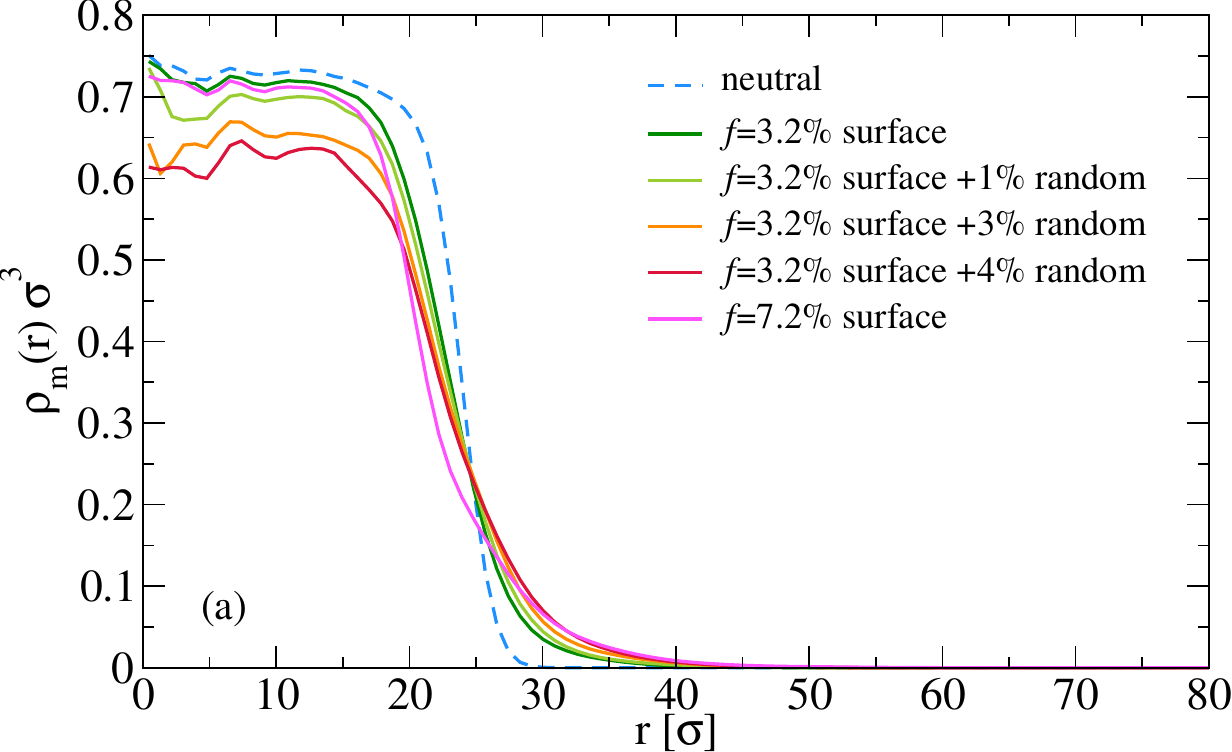}
    \hfill
    \includegraphics[width=0.47\textwidth]{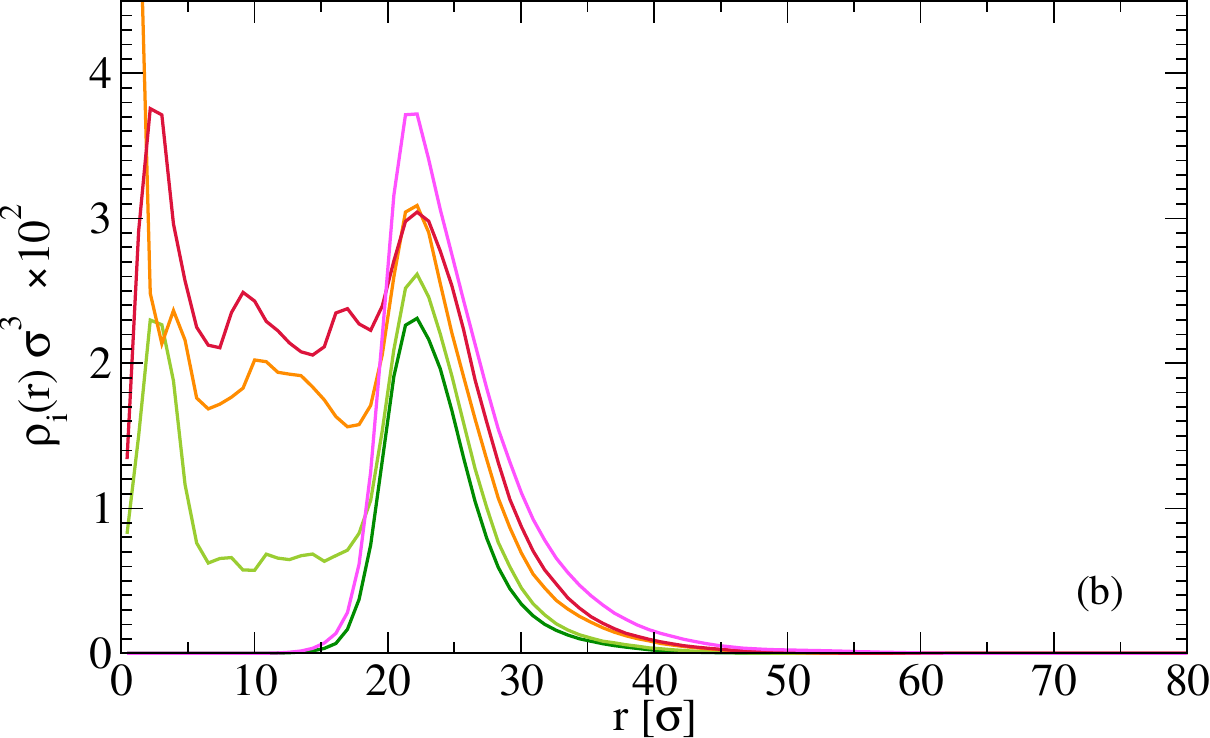}\\
    \vspace{0.5cm}
    \includegraphics[width=0.47\textwidth]{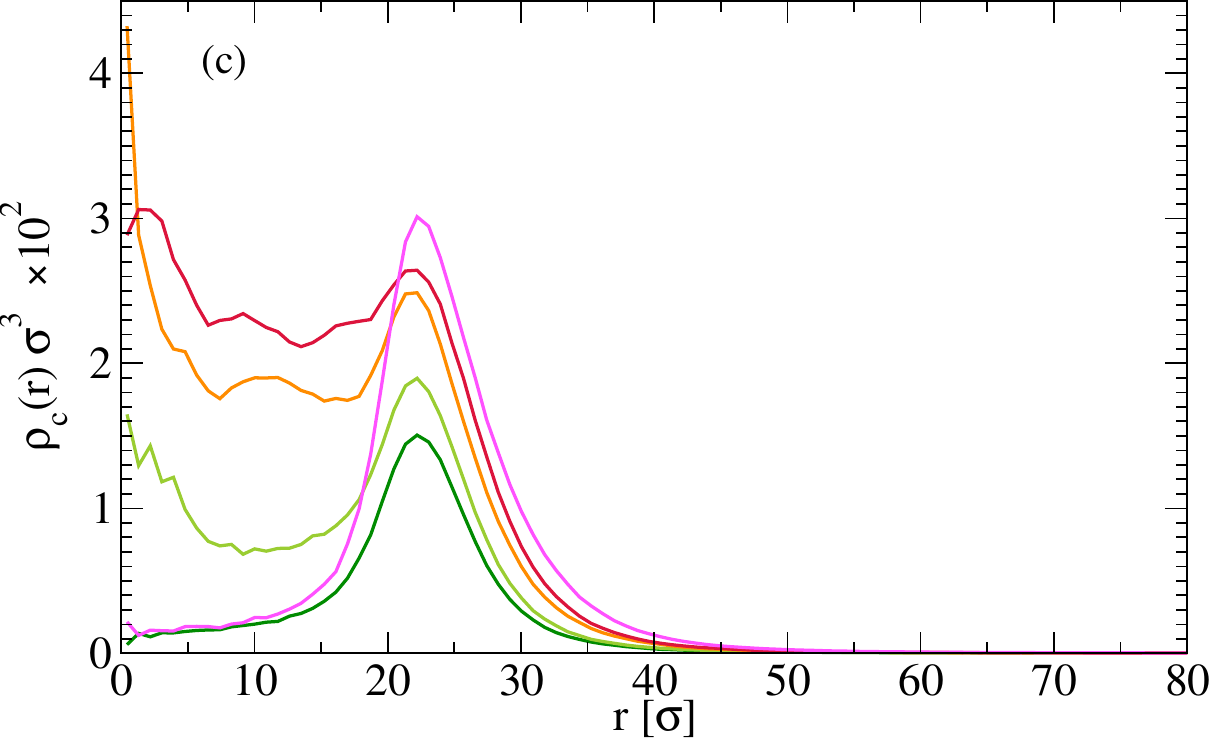}
    \hfill
    \includegraphics[width=0.47\textwidth]{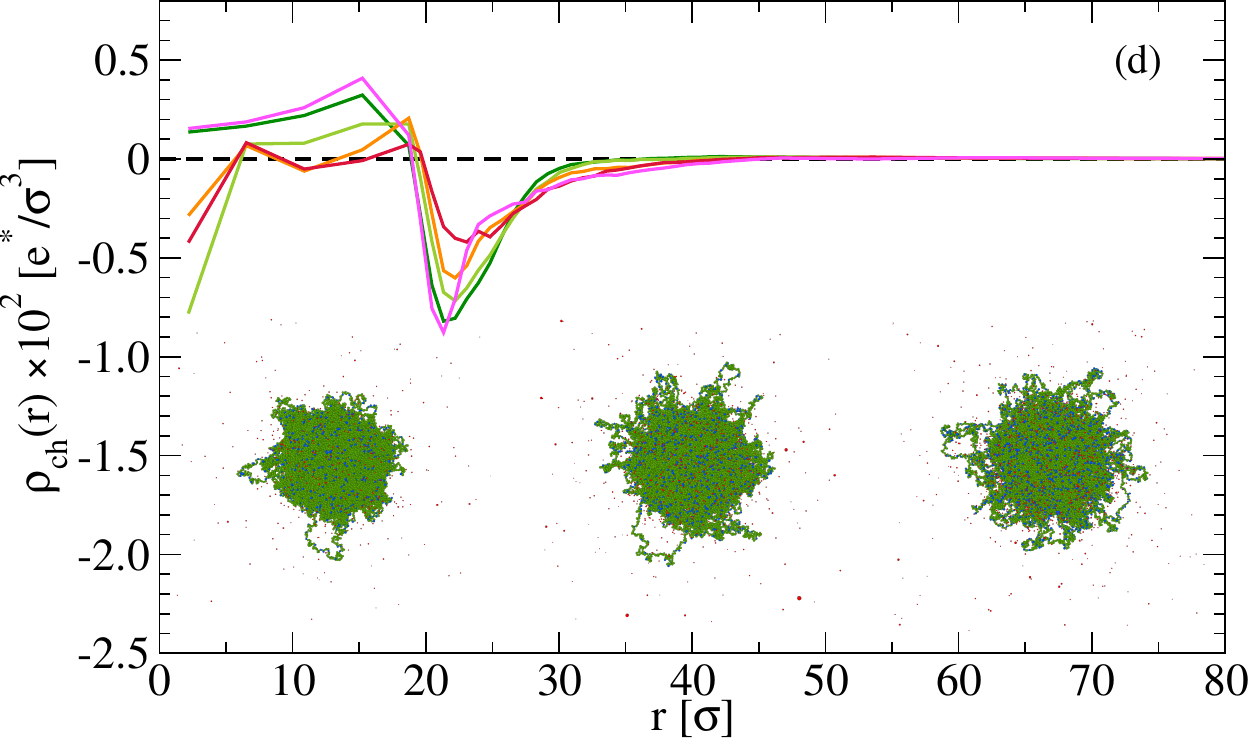}
    \caption{(a) Monomers, (b) ions, (c) counterions and (d) net charge radial density profiles for the microgels as a function of the distance from the microgel center of mass $r$ for $\alpha=0.8$. Inset in (d): snapshots corresponding to the microgels with surface charge $f_s=3.2\%$, with surface charge $f_s=3.2\%$ and random charge $f_r=4\%$, and with surface charge $f=7.2\%$, from left to right.}
    \label{fig:sim:structure:a0.80}
\end{figure}

Turning to the ion density distributions (panels (b)), we find that for all temperatures they are always peaked at the microgel periphery, 
where most of the charges are located, while a plateau at short distances is found to increase with the amount of randomly distributed beads.%, progressively entering the core region. 
The peak height, as expected, increases when microgels become more hydrophobic, since the average inter-charge distance decreases, thus producing a net increment of the charge density of the microgels (panels (d)). Such an augmented charge density has been already widely discussed in experiments~\cite{pelton_particle_1989,truzzolillo_overcharging_2018,sennato_double-faced_2021}.
The cases of microgels with only surface charges display a different behavior, where a net positive charge develops within the core of the particles due to the accumulation of counterions (panels (c)). Such a trend is not present when random charges are added, so that the net charge inside the core tends to zero, as visible at not too small distances from the centre of mass (where statistical noise becomes rather large for $r\rightarrow 0$).
In addition the net charge distribution (panels (d)) displays the largest (negative) peaks for the microgels with surface charges only. This is due to the nearly zero electric field in the microgel interior when all the charged beads are confined in its periphery. Therefore counterions are still able to penetrate and diffuse within the microgel, tending to repel further counterions that would neutralize the fixed charge distribution carried by the chains. When charges are  randomly added throughout the network, this repulsive electric field decreases, and the surface peak of negative beads increases, allowing counterions to better screen them even in the corona, resulting in a decrease of the peaks in the net charge.
%\gdst{This is evidenced by the net charge overshoots close to the main peaks, especially in collapsed microgels (Figure \ref{fig:sim:structure:a1.20}) and, since the microgel is overall electroneutral, it gives rise to an increased (net) charge density peak in the peripheral corona.}
%\DT{check the following sentence, that I modified but it can be/should improved: The fact that...}
The fact that charged beads are spatially constrained by network bonds, while counterions are free to move, generates an asymmetry resulting in a net peak of the counterion density, that is always more smeared with respect to that of charged monomers. This is also evidenced by the net charge overshoots close to the main peaks, especially in collapsed microgels (Figure \ref{fig:sim:structure:a1.20}).

%Finally, %\gdst{the resulting structure opening due to added charges are visible for the three solphofobic parameter presented here: snapshots of equilibrium configurations show that introducing random charges and confining the latter at the microgel surface progressively enhance the}
Snapshots of equilibrium configurations of the microgels with $f=3.2\%$ charges on the surface, with added $f_r=4\%$ charges randomly distributed, and with $f=7.2\%$ surface charges are also shown in panels (d). They allow us to visualize the fact that the increase of the random charges induce a larger structural inhomogeneity, a feature that is further enhanced by moving all fixed charges (monomers) within the corona. Hence, the microgel progressively acquires a more open structure and a larger amount of dangling chains that contribute to the net increase of the hydrodynamic size of the particle.  
All the discussed features are maintained as $\alpha$ increases and it is particularly evident that, even well above the VPT temperature, the more charged the microgel and the more asymmetric its charge distribution, the higher is the degree of anisotropy of the collapsed state, which maintains residual swollen chains.%\DT{not clear, you might want to say swollen chains instead of unswollen groups?}. 

This explains why a further increase of $f$, especially if charges are only distributed within the corona region, will eventually prevent the full microgel collapse or even the occurrence of the VPT in the usual range of investigated temperatures in experiments, as found for the AC5 microgels in Fig.~3  and in previous works~\cite{shibayama_dynamic_1996,kratz_influence_2000,farooqi_stability_2017}.%\ez{citare qualcosa. La chiuderei qui.  La discussione sembrava un po' incentrata su alpha=0, ho provato un pochino as armonizzarla sui tre alpha, magari va ancora aggiustata in fase di rilettura.}

%\DT{Emanuela e Giovanni: how do you want to conclude this section?}
%\gdm{Non so se potremmo dire altro di utile da comparare con gli esperimenti\ldots Io non mi dilungherei troppo oltre sulle sole simulazioni, oltre a commentare i dati che presentiamo e richiamare le caratteristiche degli esperimenti, per non ripetere cose che abbiamo gi\'a detto altrove e lasciare il discorso pi\'u breve e fluido\ldots}

\begin{figure}[ht]
 \centering
    \includegraphics[width=0.47\textwidth]{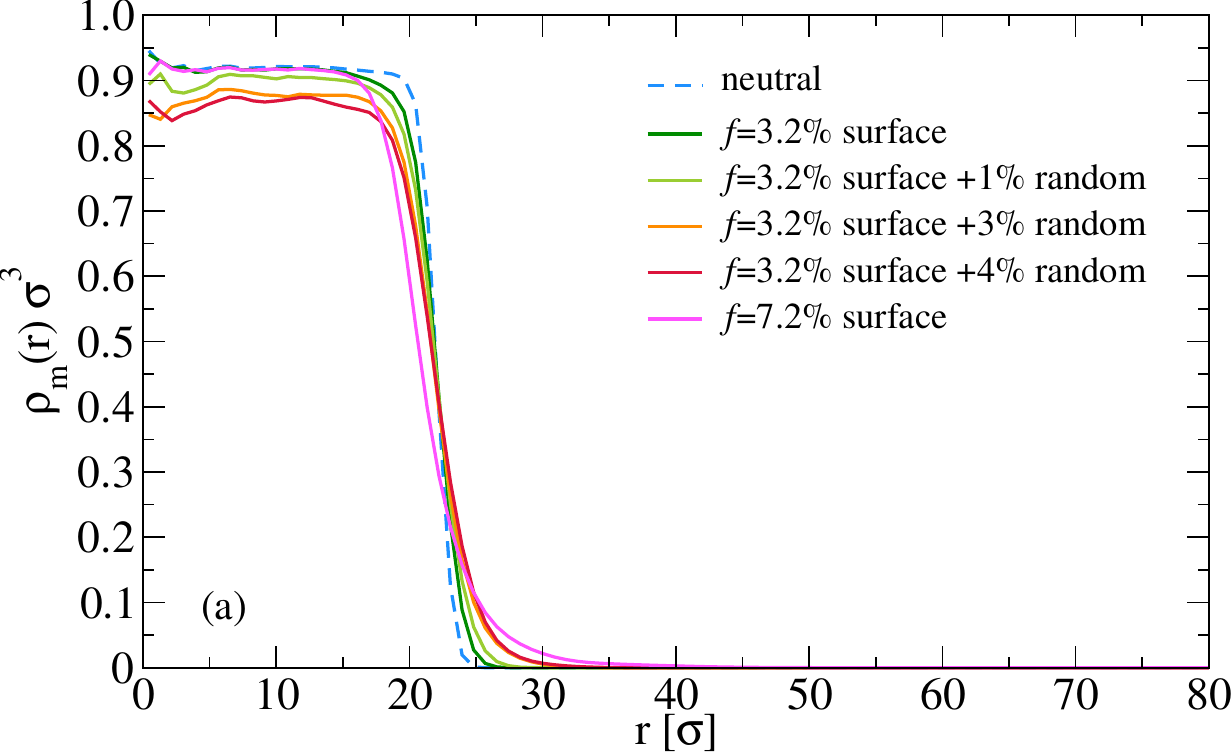}
    \hfill
    \includegraphics[width=0.47\textwidth]{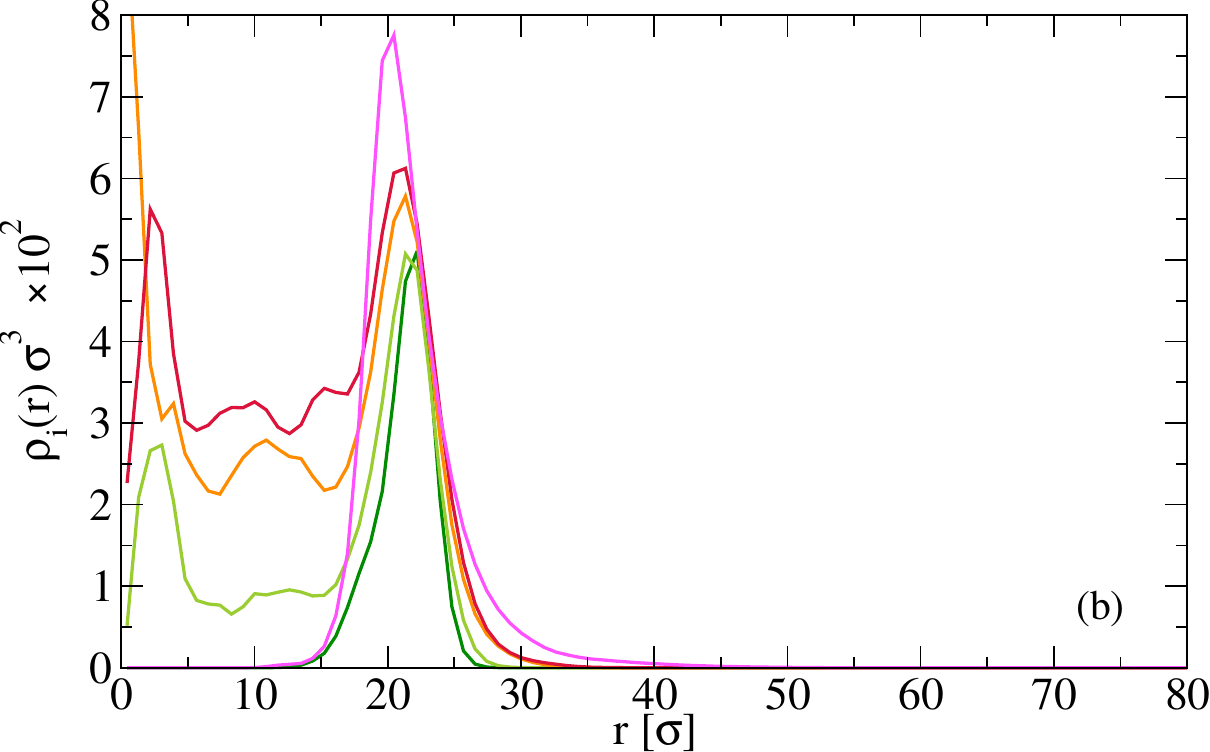}\\
    \vspace{0.5cm}
    \includegraphics[width=0.47\textwidth]{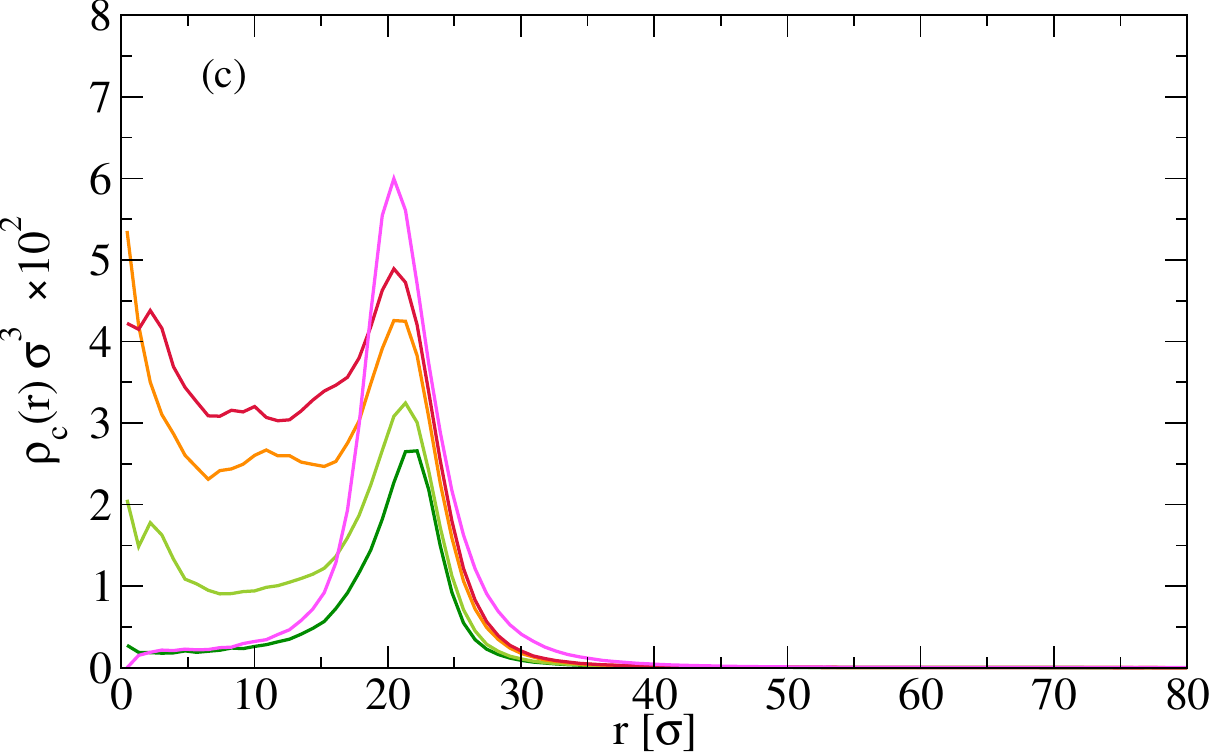}
    \hfill
    \includegraphics[width=0.47\textwidth]{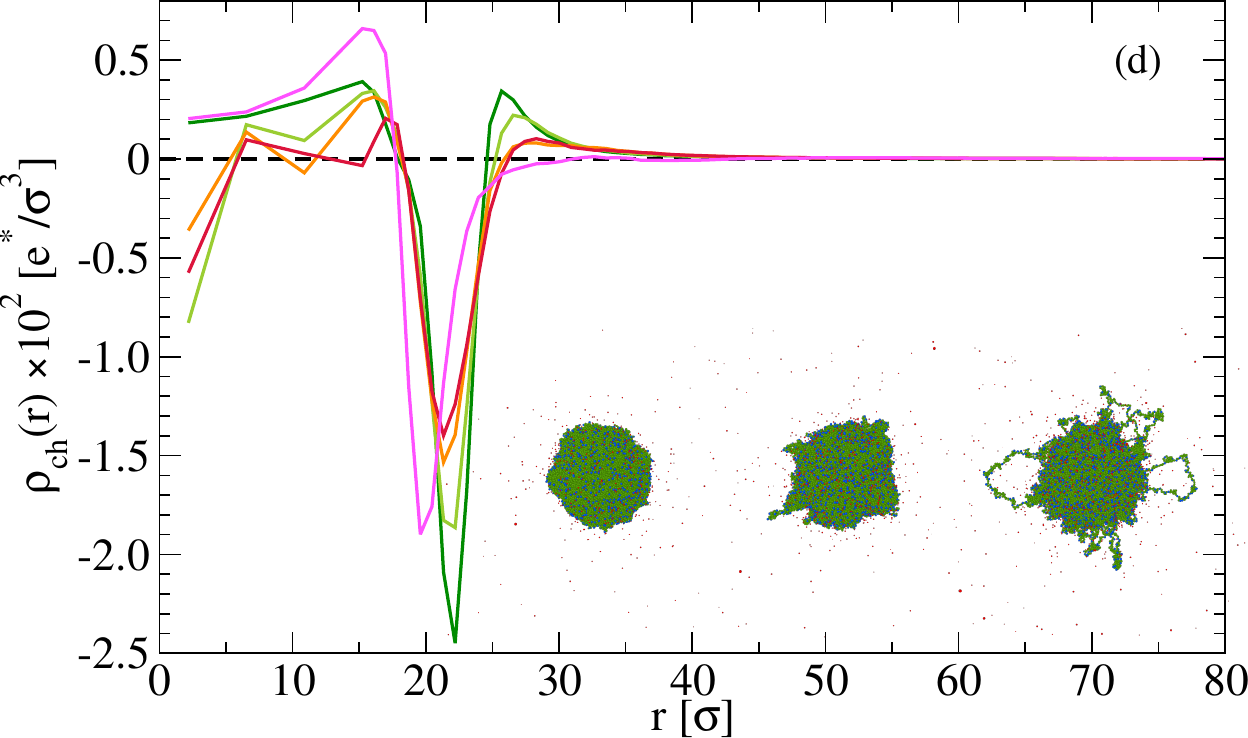}
    \caption{(a) Monomers, (b) ions, (c) counterions and (d) net charge radial density profiles for the microgels as a function of the distance from the microgel center of mass $r$ for $\alpha=1.2$. Inset in (d): snapshots corresponding to the microgels with surface charge $f_s=3.2\%$, with surface charge $f_s=3.2\%$ and random charge $f_r=4\%$, and with surface charge $f=7.2\%$, from left to right.}
    \label{fig:sim:structure:a1.20}
\end{figure}

\section{Conclusions}\label{conclusion}
By means of experiments and numerical simulations we have shown that the addition of acrylic acid monomers in PNIPAm microgels via random co-polymerization results in a large accentuation of the two-step deswelling previously reported for pure PNIPAm microgels only~\cite{del_monte_two-step_2021}. Experimentally, the AAc addition is found to i) increase microgel mass, ii) shift, smooth out and eventually suppress the occurrence of the VPT and iii) favour the thermal decoupling between the collapse of the microgel core and that of its outer corona. Simulation results fully confirm this scenario, further highlighting the crucial role of the charge distribution within the network. Indeed, they demonstrate that the two-step deswelling is enhanced by the charge preferential location in the outer corona, with a non-trivial net charge distribution arising from strongly asymmetric counterions spatial distributions. To sum up, the (random or peripheral) addition of charged beads promotes the formation of microgels with more open structures, with the outer region being more affected than the inner part of the networks. This results in an increase of the hydrodynamic size of the microgels that is more pronounced as compared to the increase of the corresponding gyration radius. Such an effect, present in swollen microgels, reflects in their thermal behavior: their external, looser and more charged corona regions collapse progressively at higher temperatures than the corresponding cores, for increasing fraction of charged monomers. 

Although our work sheds light on the effect of AAc co-polymerization on the two-step deswelling of PNIPAm microgels, many questions still remain open. In particular, a direct proof of charge location in experiments would be very valuable. To this purpose, super-resolution microscopy experiments \cite{conley_relationship_2019} could be helpful, when combined with selective labeling of the initiators and/or of charged co-monomers. Additionally, one could rely on neutron scattering measurements, where differentiation of the monomer components can be achieved via contrast variation \cite{mohanty_interpenetration_2017, scotti_hollow_2018}.

Another important question concerns whether larger charge contents or pH values would trigger different kinds of transitions (e.g. discontinuous or multi-step), as observed in previous works \cite{kratz_influence_2000}. To this aim, it would be interesting to assess how this will affect the two-step deswelling mechanism discussed in the present work. Hence, for such very charged microgels, a simultaneous measurement of $Rg$ and $R_H$ would again be very valuable. 

Finally, the outcomes of this work are relevant for future investigations of the collective behavior and dynamics of PNIPAm-based microgels, especially at high temperatures and for near-critical conditions, where numerical/theoretical works are still lacking. In fact, though the effective potential for swollen microgels can be described as a soft repulsion \cite{bergman_new_2018}, at high temperatures both hydrophobic and electrostatic interactions should become important \cite{howe_anomalous_2009,chaudhary_linear_2021} . It will be therefore necessary to appropriately assess their relative contribution to the collective and rheological behavior of dense microgel suspensions. Last but not least, polyelectrolyte and/or nanoparticle adsorption onto soft colloids, such as microgels, is very often determined by electrostatic forces that are tightly related to the swelling state of the adsorbing substrate \cite{truzzolillo_overcharging_2018,sennato_double-faced_2021}. Understanding how randomly distributed charges and the enhanced two-step deswelling influence the adsorption of oppositely charged objects onto soft ionic microgels represents another important open question having potential important implications in applications such as water remediation protocols and sensors.     
\begin{acknowledgement}
%Please use ``The authors thank \ldots'' rather than ``The
%authors would like to thank \ldots''.
We acknowledge financial support from the Agence Nationale de la Recherche (Grant ANR-20-CE06-0030-01; THELECTRA). DT thanks Luca Cipelletti for illuminating discussions.
\end{acknowledgement}

\begin{suppinfo}
We report: i) the mass of each compound introduced in all the free radical polymerization syntheses; ii) the best fits (Equation \ref{eq:critical fit}) obtained for both $R_H(T)$ and $R_g(T)$ of all the microgels showing a VPT (AC0,AC1,AC3); iii) the polydispersity index $\gamma$ obtained via cumulant analysis as detailed in section \ref{matmeth} for all the samples and temperatures.  
\end{suppinfo}

\bibliography{achemso-demo}

\end{document}

\subsection{Synthesis}
We report here the mass of the compounds dissolved in deionized water during the synthesis of PNIPAm-co-AAc microgels for different fraction of acidic monomers $f_{AAc}$ including: monomers (NIPAm and AAc), initiator (KPS), crossliker (BIS) and stabilizer (SDS).  

\begin{table}
  \caption{Mass content of chemicals used in 150mL of water in addition to BIS (91.96 (mg)) and KPS (29.86 (mg)) whose mass content was kept constant for all the syntheses.}
  \label{tbl:chem}
  \begin{tabular*}{\textwidth}{@{\extracolsep{\fill}}lllll}
    \hline
    Sample code &$f_{AAc}$ & NIPAM (mg) & AAc (mg) & SDS (mg)  \\
    \hline
    AC0 & 0\% & 1250.00 & 0     & 27.08\\
    AC1 & 1\% & 1237.50 & 7.96  & 26.80\\
    AC3 & 3\% & 1212.50 & 23.88 & 26.26\\
    AC5 & 5\% & 1187.50 & 39.79 & 25.72\\
    \hline
  \end{tabular*}
\end{table}

\subsection{Light scattering}

\begin{figure}[ht]
 \centering
    \includegraphics[width=18cm]{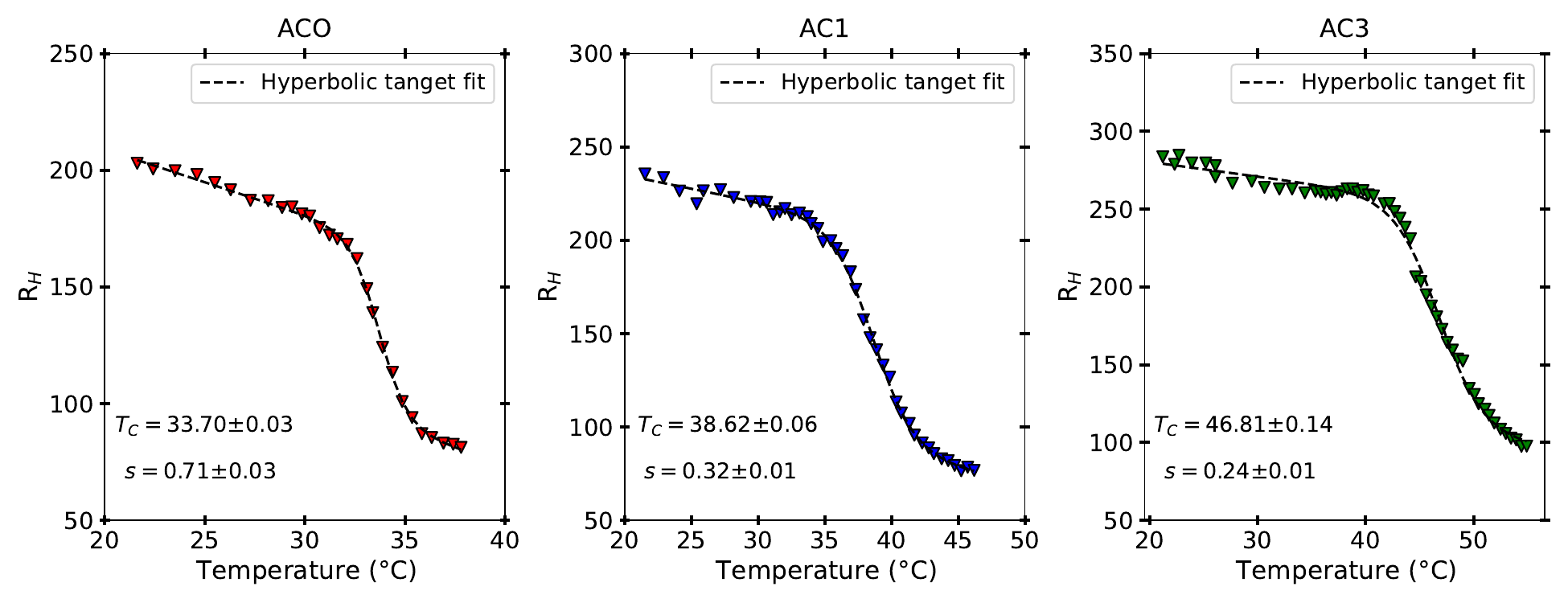}
    \caption{Hyperbolic tangent fit (Equation 15 of the main text) of $R_H(T)$ for all samples (AC0,AC1.AC3) showing a VPT.}
    \label{fig:Rh}
\end{figure}

\begin{figure}[ht]
 \centering
    \includegraphics[width=18cm]{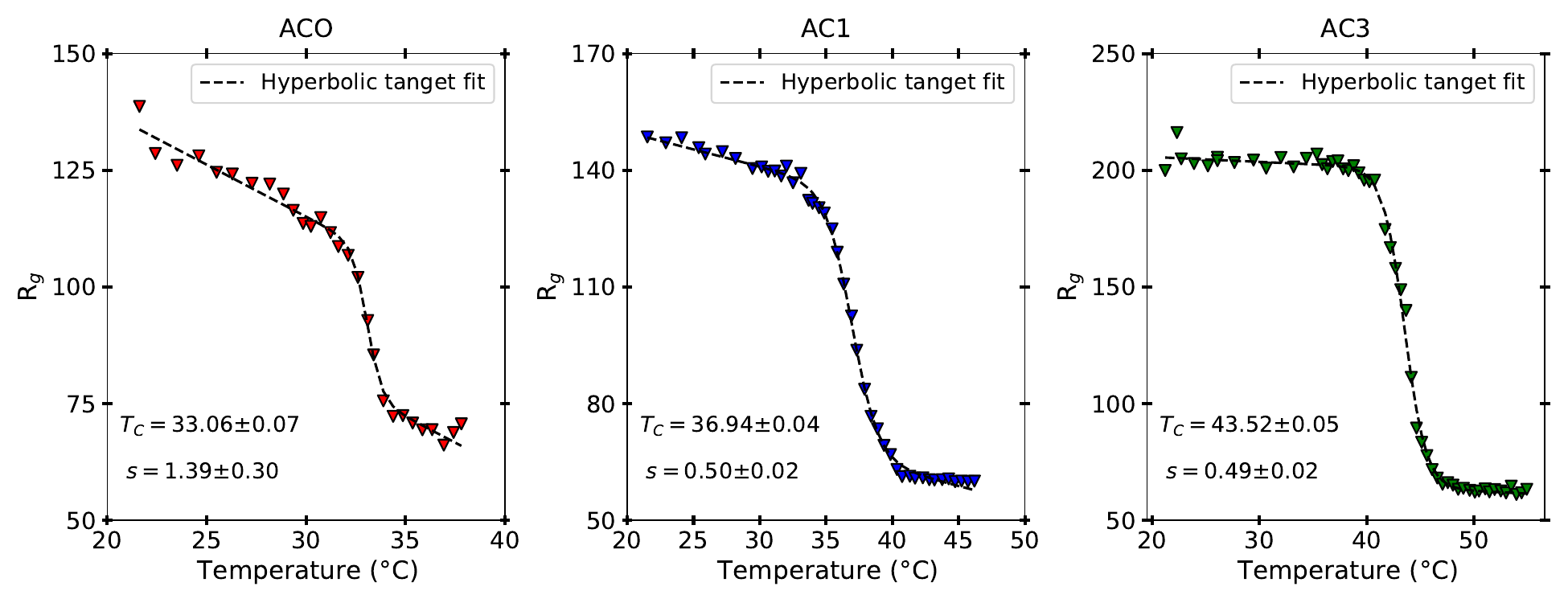}
    \caption{Hyperbolic tangent fit (Equation 15 of the main text) of $R_g(T)$ for all samples (AC0,AC1.AC3) showing a VPT.}
    \label{fig:Rg}
\end{figure}

\begin{figure}[ht]
 \centering
    \includegraphics[width=18cm]{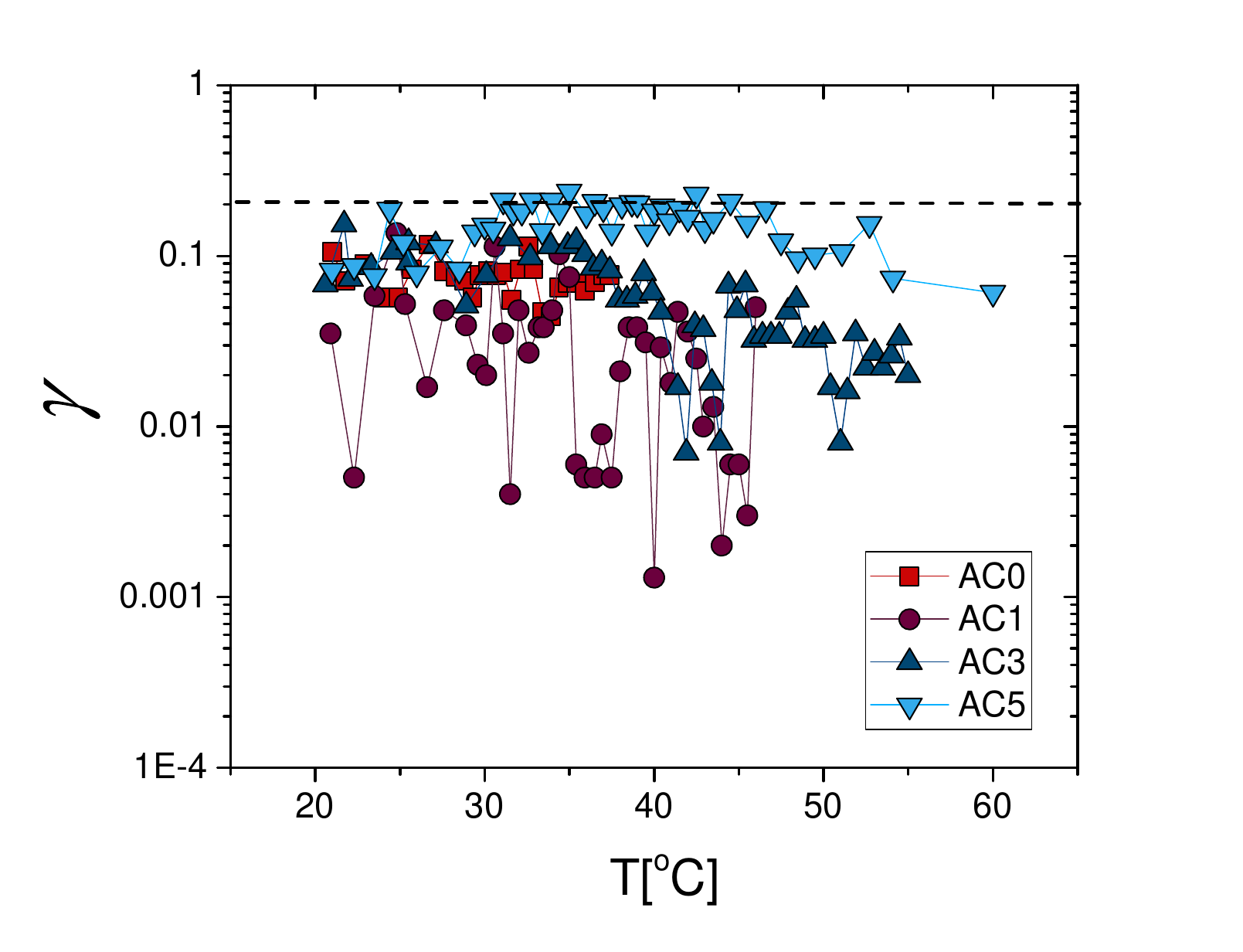}
    \caption{Polydispersity index $\gamma$ for all the investigated microgel samples as a function of temperature. The dashed horizontal line corresponds to $\gamma=0.2$}
    \label{fig:PDI}
\end{figure}

In figure \ref{fig:Rh} and \ref{fig:Rg} we report the best fit results obtained via equation 15 of the main text through which we extracted the critical temperatures $T_{H,g}$ and the "sharpness" parameters $s^{H,g}$. 
In figure \ref{fig:PDI} we report the polydispersity index $\gamma$ as defined in the main text (section 2.3). The polydispersity index never exceeds 0.23.  

